\documentclass[conference]{IEEEtran}
\IEEEoverridecommandlockouts
\usepackage{cite}
\usepackage{amsmath,amssymb,amsfonts}
\usepackage{amsthm}  % 引入定理环境支持
\newtheorem{theorem}{Theorem}         % 定理
\usepackage{graphicx}
\usepackage{textcomp}
\usepackage{xcolor}

 \usepackage{url} 
 \usepackage{listings}
 \usepackage{xcolor}
    
 \usepackage{caption}
 \usepackage{subcaption}
 \usepackage{booktabs}

 \usepackage{algorithm}
 \usepackage{algpseudocode}  % 推荐替代 algcompatible
 \makeatletter
 
 \renewcommand{\ALG@beginalgorithmic}{\normalsize}
 \makeatother

\usepackage{makecell}
\usepackage{multirow}

\def\BibTeX{{\rm B\kern-.05em{\sc i\kern-.025em b}\kern-.08em
    T\kern-.1667em\lower.7ex\hbox{E}\kern-.125emX}}
\begin{document}

\title{Robustness Assessment and Enhancement of Text Watermarking for Google's SynthID
}

% \author{\IEEEauthorblockN{Xia Han, Qi Li, Jianbing Ni} 
% \IEEEauthorblockA{Department of Electrical and Computer Engineering, Queen's University, Kingston, ON K7L 3N6, Canada}
%  Email: \{xia.h, qi.li, jianbing.ni\}@queensu.ca}
\author{\IEEEauthorblockN{Xia Han$^{*1}$, Qi Li$^{*2}$, Jianbing Ni$^{1}$ and Mohammad Zulkernine$^{2}$} 
\IEEEauthorblockA{$^1$ Department of Electrical and Computer Engineering, Queen's University, Kingston, ON K7L 3N6, Canada\\
$^2$ School of Computing, Queen's University, Kingston, ON K7L 3N6, Canada\\
Email: \{xia.h, qi.li, jianbing.ni, mz\}@queensu.ca}
%\vspace{-0.25in}
\thanks{$^*$Xia Han and Qi Li contributed equally to this work.}}

% 手动添加脚注文本

\maketitle

\begin{abstract}
%=====new version=====
% The rapid proliferation of large language models (LLMs) has led to an explosion in the volume of AI-generated text, creating a pressing need for effective methods to distinguish between human-authored and machine-generated content. Among these methods, provenance tracking through watermarking has emerged as a promising solution. However, current watermarking technologies struggle to withstand malicious tampering in real-world scenarios. 
%直接开头从synthid出发，揭露鲁棒性问题
Recent advances in LLM watermarking methods such as SynthID-Text by Google DeepMind offer promising solutions for tracing the provenance of AI-generated text. However, our robustness assessment reveals that SynthID-Text is vulnerable to meaning-preserving attacks, such as paraphrasing, copy-paste modifications, and back-translation, which can significantly degrade watermark detectability. 
To address these limitations, we propose SynGuard, a hybrid framework that combines the semantic alignment strength of Semantic Invariant Robust (SIR) with the probabilistic watermarking mechanism of SynthID-Text. Our approach jointly embeds watermarks at both lexical and semantic levels, enabling robust provenance tracking while preserving the original meaning. Experimental results across multiple attack scenarios show that SynGuard improves watermark recovery by an average of 11.1\% in F1 score compared to SynthID-Text. These findings demonstrate the effectiveness of semantic-aware watermarking in resisting real-world tampering. All code, datasets, and evaluation scripts are publicly available at: \textit{\url{https://github.com/githshine/SynGuard}}.
\end{abstract}


\section{Introduction}
Text watermarking has emerged as a promising solution for tracing the origin of AI-generated content, offering a lightweight, model-agnostic method for content provenance verification \cite{kirchenbauer2023watermark, crothers2023machine}. It identifies generated text from surface form alone, without access to the original prompt or underlying model. This makes watermarking especially appealing in open-world scenarios, where black-box models and unknown sources proliferate.

Among existing approaches, Google DeepMind’s SynthID-Text is state-of-the-art \cite{dathathri2024scalable}, notable as the only watermarking method integrated into a real-world product (Google’s Gemini models), a rare industrial deployment in this domain. It embeds imperceptible statistical signals during generation via tournament sampling, departing from earlier post-hoc or green-list based methods \cite{liu2024survey,kirchenbauer2023watermark}. This approach introduces controlled stochasticity in token selection and shows improved detectability in benign settings. However, its resilience to malicious tampering remains underexplored. Previous studies note the fragility of lexical watermarks under meaning-preserving, surface-altering transformations \cite{wang-etal-2025-morphmark,liu2024semanticinvariantrobustwatermark}; SynthID-Text, despite advancements, shares this limitation, motivating deeper analysis of its practical robustness.  

In this work, we systematically assess SynthID-Text under real-world meaning-preserving transformations: paraphrasing, synonym substitution, copy-paste rearrangement, and back-translation, attacks preserving semantic content while modifying lexical or syntactic surface form. Results reveal a critical vulnerability: detection accuracy drops sharply even under light paraphrasing or translation. These findings align with prior concerns, highlighting a gap in current capabilities. 

To address this, we propose \textbf{SynGuard}, a hybrid scheme integrating Semantic Invariant Robust (SIR) alignment \cite{liu2024semanticinvariantrobustwatermark} with SynthID’s token-level probabilistic masking. Our method embeds provenance signals at both lexical and semantic levels: the semantic component guides generation toward SIR-favored contexts (enhancing robustness to synonym and paraphrase attacks), while SynthID’s token logic retains seed-derived randomness (resisting keyless removal).  

Unlike prior lexical-only approaches \cite{kirchenbauer2023watermark,dathathri2024scalable}, SynGuard adds a semantic signal to detect tampering that preserves meaning but alters surface structure. This hybrid design better balances false positive rate and tampering robustness. We formalize this via theoretical analysis (Section~\ref{sec:robustness_analysis}), showing semantically consistent transformations rarely suppress SIR-guided scores unless meaning is significantly distorted, one of the first formal analyses of watermark resilience under semantic equivalence.  

Empirical evaluation across four attacks shows SynGuard improves average F1 by \textbf{11.1\%} over SynthID-Text, performing especially well under paraphrasing and round-trip translation (common in content reposting and cross-lingual reuse). We uncover a new vulnerability axis: back-translation-induced watermark degradation correlates with translation quality, as poorer machine translation distorts signals more even with preserved semantics. This insight introduces new considerations for evaluating robustness across linguistic contexts and highlights the need for multilingual benchmarks.  

Our contributions are summarized as follows:  
\begin{enumerate}
    \item Conduct the first comprehensive robustness evaluation of SynthID-Text under four meaning-preserving transformations: paraphrasing, synonym substitution, copy-paste tampering, back-translation.  
    \item Propose SynGuard, a hybrid algorithm combining semantic-aware token preferences with token-level probabilistic sampling.  
    \item Demonstrate SynGuard consistently improves detection robustness, particularly for surface-altered but meaning-preserved content.  
    \item Reveal back-translation attack vulnerability correlates with machine translation quality, an overlooked axis.  
\end{enumerate}

\section{Related Work}

% Text watermarking is a technique for distinguishing AI-generated text from human-written content by embedding specific information into text sequences without compromising their quality. According to the point in the text generation pipeline when the watermark is inserted, watermarking methods can be categorized into two types~\cite{liu2024survey}: watermarking for existing text and watermarking during text generation.
% The first type, watermarking for existing text, adds watermarks through post-processing of pre-existing text. This is typically done by reformatting sentence structures using different Unicode characters, or by altering lexical choices or syntactic constructions. While these methods are relatively easy to implement, they are also relatively easy to remove through text reformatting or normalization.
Text watermarking distinguishes AI vs human text by embedding specific information into text sequences without quality loss. By watermark insertion stage in text generation, methods fall into two types~\cite{liu2024survey}: watermarking for existing text and during generation. The first type adds watermarks via post-processing of existing text, typically via reformatting sentences with Unicode, altering lexicon or syntax. Though easy to implement, they are easy to remove via reformatting/normalization.

Watermarking during generation is achieved by modifying logits in token generation. This approach is more stable, imperceptible, harder for attackers to detect/remove. A key method is the KGW algorithm \cite{kirchenbauer2023watermark}: it splits vocabulary into green/red lists via pseudorandom seed. Adding positive bias to green list tokens makes them more likely selected than red ones. This skew enables high-confidence post hoc detection. KGW balances robustness and imperceptibility, underpinning recent frameworks~\cite{zhao2023provablerobustwatermarkingaigenerated, hu2023unbiasedwatermarklargelanguage, kirchenbauer2023reliability}.
% On the other hand, watermarking during text generation is usually achieved by modifying the logits during the token generation process. This approach tends to be more stable and imperceptible, making it more difficult for attackers to detect or remove. One of the most influential generation-time watermarking methods is the KGW algorithm \cite{kirchenbauer2023watermark}. It divides the vocabulary into two lists, a green list and a red list, based on a pseudorandom seed. By adding a positive bias to tokens in the green list, these tokens tend to be selected during generation than those in the red list. This statistical skew enables post hoc detection of watermarked text with high confidence. The KGW method strikes a balance between robustness and imperceptibility and forms the basis for several recent watermarking frameworks~\cite{zhao2023provablerobustwatermarkingaigenerated, hu2023unbiasedwatermarklargelanguage, kirchenbauer2023reliability}.

Google DeepMind's SynthID-Text~\cite{dathathri2024scalable} advances generation-based watermarking by using pseudorandom functions (PRFs) and tournament sampling to guide token generation in a more randomized and less perceptible manner. 
During the sampling process, each token candidate is assigned $m$ independent $g$-values $(g_1, ..., g_m)$, and the token with the highest total $g$-value (e.g., the sum of all $g_i$) among all candidates is selected. 
These $g$-values can later be used for watermark detection. This design improves robustness against removal attacks such as truncation and basic paraphrasing.

Despite these strengths, most generation-time watermarking algorithms, including SynthID-Text, do not incorporate semantic information when adjusting logits. As a result, they remain vulnerable to semantic-preserving adversarial attacks. Recent studies have begun exploring semantic-aware watermarking strategies~\cite{liu2024semanticinvariantrobustwatermark,ren2023robust,he2024can}.
A Semantic Invariant Robust watermarking algorithm is introduced \cite{liu2024semanticinvariantrobustwatermark}, which maps extracted semantic features from preceding context into the logit space to guide next-token generation. In this approach, semantic similarity becomes a key indicator for detecting watermarks. While promising in terms of robustness, this method relies on additional language models, which increases computational complexity and resource consumption. Furthermore, enforcing semantic consistency reduces output diversity and naturalness.

% In this work, we evaluate the robustness of SynthID-Text under a range of text editing attacks. Based on these results, we further explore a hybrid watermarking mechanism that integrates token-level and semantic-level signals to enhance resistance against semantic-preserving transformations.

\section{Preliminaries}

% This section introduces the background and basic concepts relevant to this topic. Section~\ref{3.1} explains how a large language model (LLM) functions and how it selects the next token during the text generation process. Section~\ref{3.2} describes various approaches to embedding watermarks into text, and Section~\ref{3.3} discusses the key challenges in text watermarking.

\subsection{Large Language Model}
\label{3.1}
A large language model (LLM) $M$ operates over a defined set of tokens, known as the vocabulary $V$. Given a sequence of tokens $t = [t_0, t_1, \ldots, t_{T-1}]$, also referred to as the \textit{prompt}, the model computes the probability distribution over the next token $t_T$ as $P_M(t_T \mid t_{:T-1})$.
The model $M$ then samples one token from the vocabulary $V$ according to this distribution and other sampling parameters (e.g., temperature). This process is repeated iteratively until the maximum token length is reached or an end-of-sequence (EOS) token is generated.

This next-token prediction is typically implemented using a neural network architecture called the Transformer~\cite{vaswani2017attention}. The process involves two main steps:
\begin{enumerate}
    \item The Transformer computes a vector of logits $z_T = M_{t_{:T-1}}$ over all tokens in $V$, based on the current context $t_{:T-1}$.
    \item The softmax function is applied to these logits to produce a normalized probability distribution: $P_M(t_T \mid t_{:T-1})$.
\end{enumerate}

%该部分要删除简化，至少缩掉一半篇幅 
%\subsection{Text Watermarking for LLMs}
\subsection{SynthID-Text in LLM Text Watermarking}
\label{3.2}
Text watermarking for LLMs operates mainly at two stages: embedding-level (modifying internal embedding vectors, which is complex and less generalizable) and generation-level (altering token generation via logits adjustment or sampling strategies). Generation-level methods include logits-based approaches (e.g., KGW algorithm \cite{kirchenbauer2023watermark}, biasing logits toward ``green list" tokens) and sampling-based approaches (e.g., Christ algorithm \cite{christ2023undetectablewatermarkslanguagemodels}, using pseudorandom functions to guide sampling without logit modification).

SynthID-Text is a sampling-based algorithm featuring a novel tournament sampling mechanism for token selection. Candidate tokens are sampled from the original LLM-generated probability distribution \( p_{LM} \), so higher-probability tokens may appear multiple times in the candidate set. Each candidate token is evaluated using \( m \) independent pseudorandom binary watermark functions \( g_1, g_2, ..., g_m \). These functions assign a value of 0 or 1 to a token \( x \in V \) based on both the token and a random seed \( r \in \mathbb{R} \): $g_l(x, r) \in \{0, 1\}.$ The tournament sampling procedure selects the token with statistically high \( g \)-values across the \( m \) functions, while respecting the base LLM distribution. To detect if a text \( t = [t_1, ..., t_T] \) is watermarked, the average \( g \)-value across all tokens and functions is computed:
\begin{equation}
\text{Score}(t) = \frac{1}{mT} \sum_{i=1}^{T} \sum_{l=1}^{m} g_l(t_i, r_i).
\end{equation}

\subsection{Text Watermarking Challenges}
\label{3.3}

Compared to watermarking techniques in other media such as images or audio~\cite{chen2019deepmarks,qiao2023novel,zhang2021deep,darvish2019deepsigns}, embedding watermarks in text introduces a distinct set of challenges:

\textbf{Token Budget Constraints:} A standard $256 \times 256$ image offers over 65K potential pixel positions for embedding watermarks~\cite{neekhara2022facesigns}. In contrast, the maximum token length for LLMs like GPT-4 is around 8.2K tokens (with limited access to 32K\footnote{\url{https://openai.com/index/gpt-4-research/}}), which is significantly smaller. This limited capacity makes it harder to embed watermarks without detection by human readers and increases vulnerability to adversarial edits. As a result, watermarking algorithms for text require more careful design to ensure both imperceptibility and robustness.

\textbf{Perturbation Sensitivity:} Text data is highly sensitive to editing~\cite{zhao2023protecting}. While small pixel changes in an image are often imperceptible to the human eye, even minor alterations in a text, such as character replacements or word substitutions, can be easily noticed by readers or detected by spelling and grammar tools. Moreover, replacing entire words can unintentionally alter the meaning, introduce ambiguity, or degrade sentence fluency.

\textbf{Vulnerability:} Watermarks in text are particularly susceptible to removal through common natural language transformations. An attacker can easily re-edit the content by substituting synonyms, or paraphrasing with new sentence structures~\cite{qiu2022adversarial}.

% \subsection{Equations}
% Number equations consecutively. To make your 
% equations more compact, you may use the solidus (~/~), the exp function, or 
% appropriate exponents. Italicize Roman symbols for quantities and variables, 
% but not Greek symbols. Use a long dash rather than a hyphen for a minus 
% sign. Punctuate equations with commas or periods when they are part of a 
% sentence, as in:
% \begin{equation}
% a+b=\gamma\label{eq}
% \end{equation}

% Be sure that the 
% symbols in your equation have been defined before or immediately following 
% the equation. Use ``\eqref{eq}'', not ``Eq.~\eqref{eq}'' or ``equation \eqref{eq}'', except at 
% the beginning of a sentence: ``Equation \eqref{eq} is . . .''

% \subsection{Figures and Tables}
% \paragraph{Positioning Figures and Tables} Place figures and tables at the top and 
% bottom of columns. Avoid placing them in the middle of columns. Large 
% figures and tables may span across both columns. Figure captions should be 
% below the figures; table heads should appear above the tables. Insert 
% figures and tables after they are cited in the text. Use the abbreviation 
% ``Fig.~\ref{fig}'', even at the beginning of a sentence.
\section{Evaluating the Robustness of SynthID-Text}
\label{ch:synthID_results}

This chapter presents the experimental settings, evaluation metrics, and results from robustness analysis of the SynthID-Text watermarking algorithm. Section~\ref{4.1} outlines the experimental setup, including the backbone model, dataset, and metrics used for evaluation. Sections~\ref{4.2} through~\ref{4.5} report SynthID-Text's performance under four types of text editing attacks: synonym substitution, copy-and-paste, paraphrasing, and re-translation. Finally, Section~\ref{4.6} summarizes and compares results across all attack types to provide a comprehensive evaluation.

\subsection{Experimental Setup}
\label{4.1}

\indent \textbf{Backbone Model and Dataset.} All experiments were conducted using \texttt{Sheared-LLaMA-1.3B} \cite{xia2023sheared}, a model further pre-trained from \texttt{meta-llama/Llama-2-7b-hf}\footnote{\url{https://huggingface.co/meta-llama/Llama-2-7b-hf}}. The model used is publicly available via HuggingFace\footnote{\url{https://huggingface.co/princeton-nlp/Sheared-LLaMA-1.3B}}.
For the dataset, we adopt the Colossal Clean Crawled Corpus (C4)~\cite{raffel2020exploring}, which includes diverse, high-quality web text. Each C4 sample is split into two segments: the first segment serves as the prompt for generation, while the second (human-written) segment is used as reference text. These unaltered human texts are treated as control data for evaluating the watermark detector’s false positive rate.

\textbf{Evaluation Metrics.} The robustness of SynthID-Text is evaluated using the following metrics:
\begin{itemize}
    \item \textbf{True Positive Rate (TPR):} The proportion of watermarked texts correctly identified.
    \item \textbf{False Positive Rate (FPR):} The proportion of unwatermarked texts incorrectly identified as watermarked.
    \item \textbf{F1 Score:} The harmonic mean of precision and recall, computed at the best threshold.
    \item \textbf{ROC-AUC:} The area under the Receiver Operating Characteristic (ROC) curve, measuring overall classification performance across all thresholds.
\end{itemize}
Each experiment was conducted using 200 watermarked and 200 unwatermarked samples, each with a fixed length of \( T = 200 \) tokens. All experiments were implemented using the MarkLLM toolkit~\cite{pan2024markllm}.

\subsection{Synonym Substitution Attack}
\label{4.2}

Given an original text sequence, the synonym substitution attack aims to replace words with their synonyms until a specified replacement ratio $\epsilon$ is reached, or no further substitutions are possible. This approach maintains semantic fidelity while subtly altering the lexical surface of the text. A well-chosen $\epsilon$ ensures that the semantic meaning remains largely intact, which aligns with the attack's objective—to disrupt watermark detection without affecting readability or content.

In this work, synonym replacement is guided by a context-aware language model to ensure substitutions remain semantically appropriate. Specifically, we implemented a method that uses WordNet \cite{miller1995wordnet}, a widely used lexical database of English, to retrieve synonym sets for eligible words. For each target word, a synonym is randomly selected using the NumPy library’s random function \cite{harris2020array}. 
The substitution is further refined using BERT-Large \cite{devlin2019bertpretrainingdeepbidirectional}, which predicts contextually suitable replacements. The process is repeated iteratively until the desired substitution ratio $\epsilon$ is reached or no more valid substitutions remain. This ensures the altered text remains semantically coherent while maximally disrupting watermark patterns.

\textbf{Details of the BERT Span Attack.} To perform context-aware synonym substitution, BERT-Large\footnote{https://huggingface.co/google-bert/bert-large-uncased} is first used to tokenize the watermarked text. Then, eligible words are iteratively replaced with contextually appropriate synonyms until either the maximum replacement ratio $\epsilon$ is reached or no further substitutions are possible. The substitution process proceeds as follows:

\begin{itemize}
    \item Randomly select a word that has at least one synonym and replace it with a \texttt{[MASK]} token:
    \begin{lstlisting}[language=Python, caption={Word Masking}, label=list:mask_example, numberblanklines=false]
    "I love programming."
    "I [MASK] programming."
    \end{lstlisting}
    \item Feed the masked sentence into the BERT-Large model, which produces a logits vector over the vocabulary using a forward pass.
    \item Rank all candidate words based on their logits and select the word with the highest probability to replace the masked token.
\end{itemize}

BERT-Large is chosen for its bidirectional architecture, allowing it to consider both preceding and succeeding context when predicting the masked word. This contextual understanding ensures that substituted words maintain semantic consistency with the original sentence.

\begin{figure}[tb]
    \centering
    
    \begin{subfigure}[b]{0.47\linewidth}
        \includegraphics[width=\linewidth]{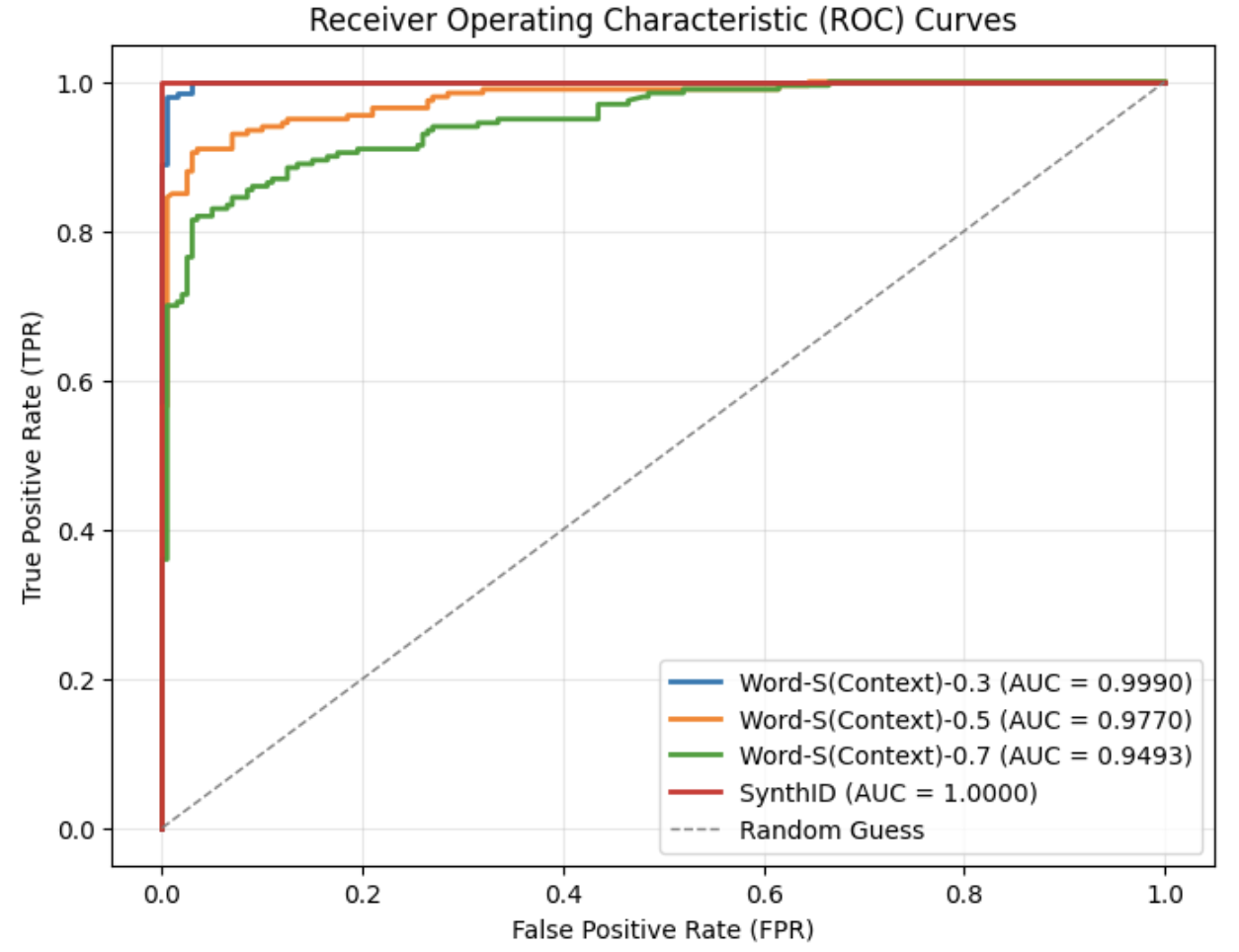}
        \caption{Overall ROC curves under synonym substitution with different replacement ratios}
        \label{fig:synonym_substitution_synthid_text}
    \end{subfigure}
    \hfill
    \begin{subfigure}[b]{0.47\linewidth}
        \includegraphics[width=\linewidth]{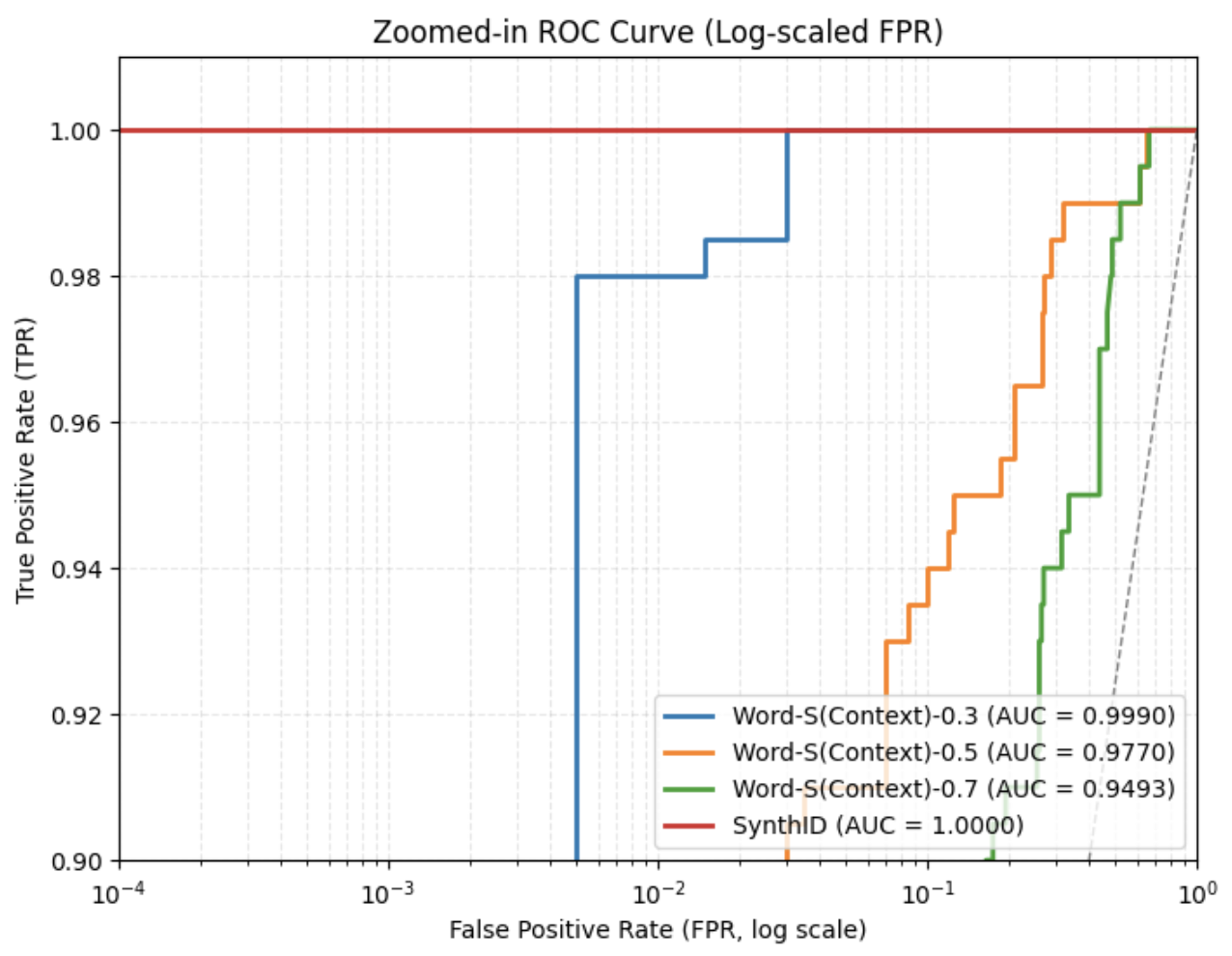}
        \caption{Zoomed-in ROC curves under synonym substitution with different replacement ratios}
        \label{fig:zoom_in_synonym_synthid}
    \end{subfigure}
     \caption{ROC curves of SynthID-Text under synonym substitution attacks with varying replacement ratios.}
    \label{fig:synthid_synonym_substitution_combined}
\end{figure}

After applying the synonym substitution strategy to a set of 200 watermarked texts, each with a token length of $T = 200$, the resulting ROC curves are presented in Fig. ~\ref{fig:synthid_synonym_substitution_combined}. As shown, the area under the curve (AUC) gradually decreases as the replacement ratio increases. Even with a replacement ratio as high as 0.7, the AUC remains above 0.94, and the corresponding F1 score is relatively high at 0.884, as reported in Table~\ref{tab:Synonym substitution_attack_summary for SynthID}. These results demonstrate that SynthID-Text exhibits strong robustness against context-preserving lexical substitutions.

\begin{table}[tb]
    \centering
    \caption{Watermark detection accuracy under different synonym substitution attack ratios.}
    \label{tab:Synonym substitution_attack_summary for SynthID}
    \begin{tabular}{lccc}     
        \toprule
        Attack               &  TPR   & FPR  &  F1 with best threshold \\
        \midrule
        No attack            &  1.0  &  0.0 &  1.0 \\
        Word-S(Context)-0.3     & 0.98  & 0.005 &  0.987 \\
        Word-S(Context)-0.5    &  0.91 & 0.035 & 0.936 \\
        Word-S(Context)-0.7    & 0.82   & 0.035 &  0.884 \\
        \bottomrule
    \end{tabular}
\end{table}

\subsection{Copy-and-Paste Attack}
\label{4.3}

Unlike synonym substitution attacks, the copy-and-paste attack does not alter the original watermarked text. Instead, it embeds the watermarked segment within a larger body of human-written or unwatermarked content. This type of attack exploits the fact that detection algorithms typically analyze text holistically; by diluting the watermarked portion, the overall watermark signal becomes weaker and harder to detect.

Prior work \cite{kirchenbauer2023reliability} has shown that when the watermarked portion comprises only 10\% of the total text, the attack can outperform many paraphrasing methods in reducing watermark detectability. In this work, we experiment with different copy-and-paste ratios and evaluate the detection performance to assess robustness.

\begin{figure}[tb]
    \centering
    \includegraphics[width=0.8\linewidth]{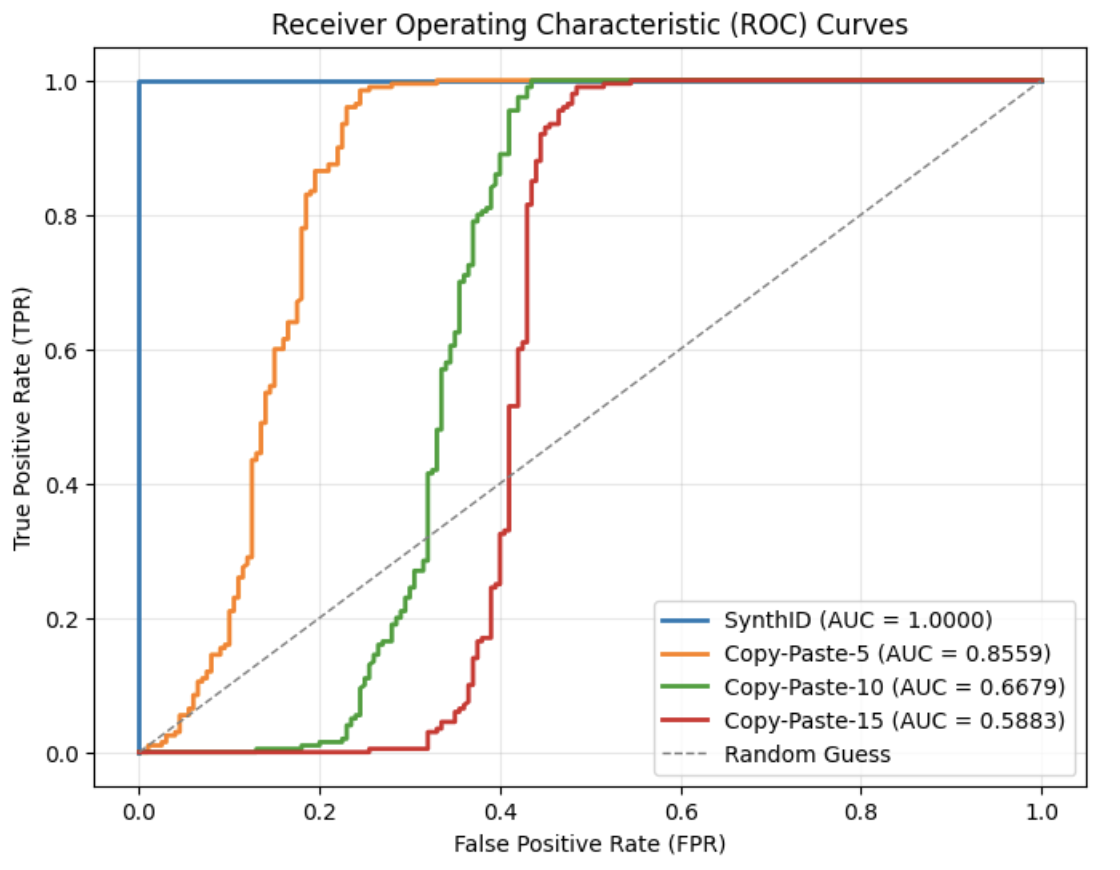}
    \caption{ROC curves under different copy-and-paste attack ratios. The blue curve represents the original SynthID-Text ROC curve without attack; the gray curve indicates random guessing. Other curves depict results under varying ratios, where the ratio denotes how many times longer the inserted natural text is compared to the original watermarked text.}

    \label{fig:copy-paste attack roc}
\end{figure}

Fig.~\ref{fig:copy-paste attack roc} presents the ROC curves for varying copy-and-paste ratios. The green curve represents the case where the added natural text is ten times longer than the original watermarked text, resulting in an AUC of 0.62—only slightly above random guess. As shown in Table~\ref{tab:copy-paste_attack_summary}, the false positive rate (FPR) for ratio $=10$ reaches 0.53, meaning that more than half of unwatermarked texts are incorrectly identified as watermarked.
As the copy-and-paste ratio increases, detection performance degrades further. When the ratio reaches 20 or higher, the AUC decreases to around or below 0.5, effectively equating to or falling below random guessing performance.

\begin{table}[tb]
    \centering
    % \caption{The watermark detection accuracy for different Copy-and-Paste attacks}
    \caption{Watermark detection accuracy under different copy-and-paste attack ratios}
    \label{tab:copy-paste_attack_summary}
    \begin{tabular}{lccc}     
        \toprule
        % \multicolumn{2}{c}{Bike} \\
        % \cmidrule(r){1-2}
        Attack               &  TPR   & FPR  &  F1 with best threshold \\
        \midrule
        No attack            &  1.0  &  0.005 &  0.9975 \\
        Copy-and-Paste-5     & 0.985  & 0.27 &  0.874 \\
        Copy-and-Paste-10    &   0.995   & \textcolor{red}{0.53} &  0.788 \\
        Copy-and-Paste-20    & 0.99   & \textcolor{red}{0.565} &  0.775 \\
        Copy-and-Paste-30    & 0.99  & \textcolor{red}{0.565} & 0.775 \\
        \bottomrule
    \end{tabular}
\end{table}

\subsection{Paraphrasing Attack}
\label{4.4}

Paraphrasing attacks aim to modify the structure and wording of a paragraph while preserving its original semantic meaning. This is typically done by rephrasing sentences or altering word choice and sentence order. Therefore, paraphrasing can be characterized along two key dimensions: \textbf{lexical diversity}, which measures variation in vocabulary, and \textbf{order diversity}, which reflects changes in sentence or phrase order.

\begin{figure}[tb]
    \centering

    \begin{subfigure}[b]{0.465\linewidth}
        \includegraphics[width=\linewidth]{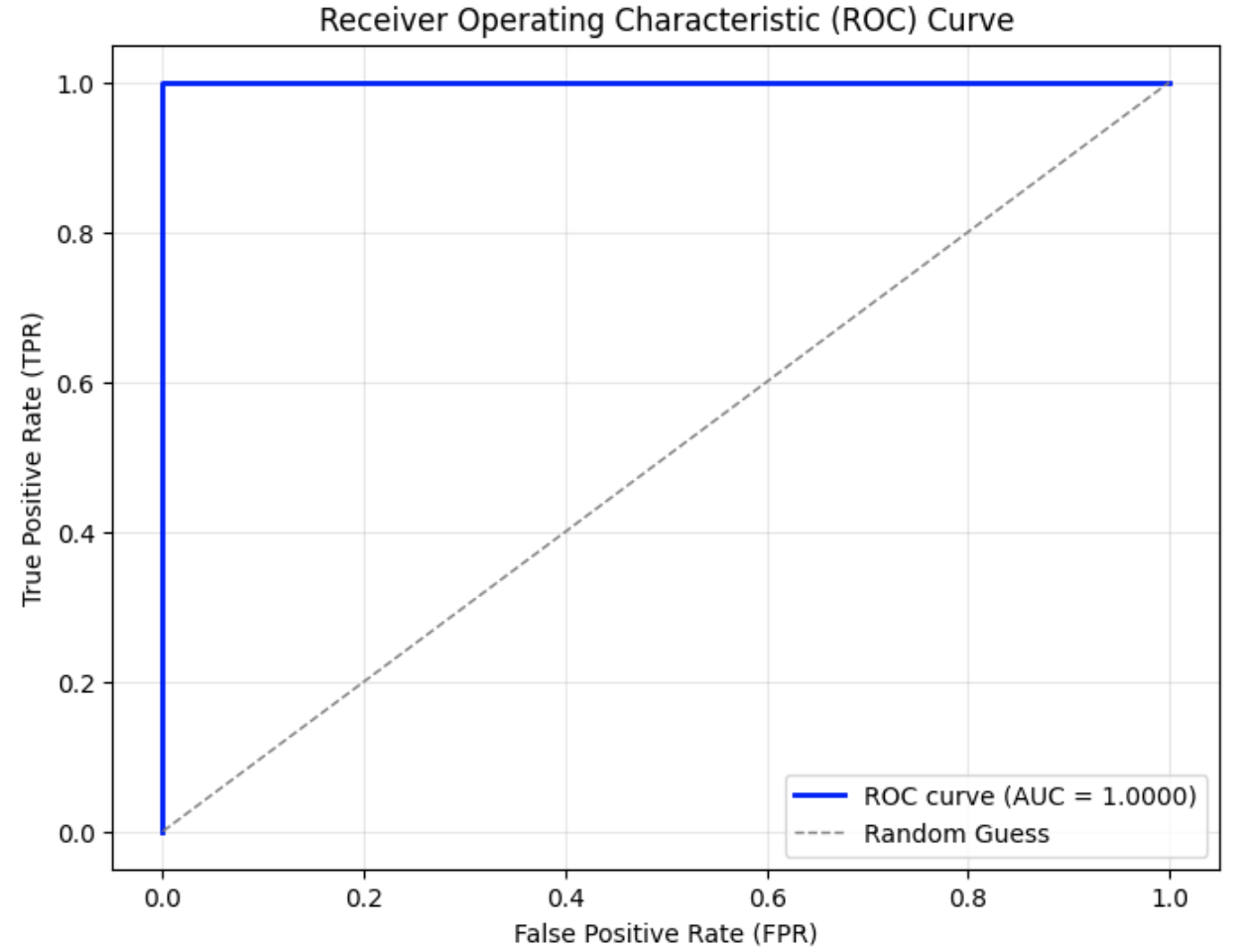}
        \caption{No attack (original SynthID-Text)}
        \label{fig:synthid_solo}
    \end{subfigure}
    \hfill
    \begin{subfigure}[b]{0.465\linewidth}
        \includegraphics[width=\linewidth]{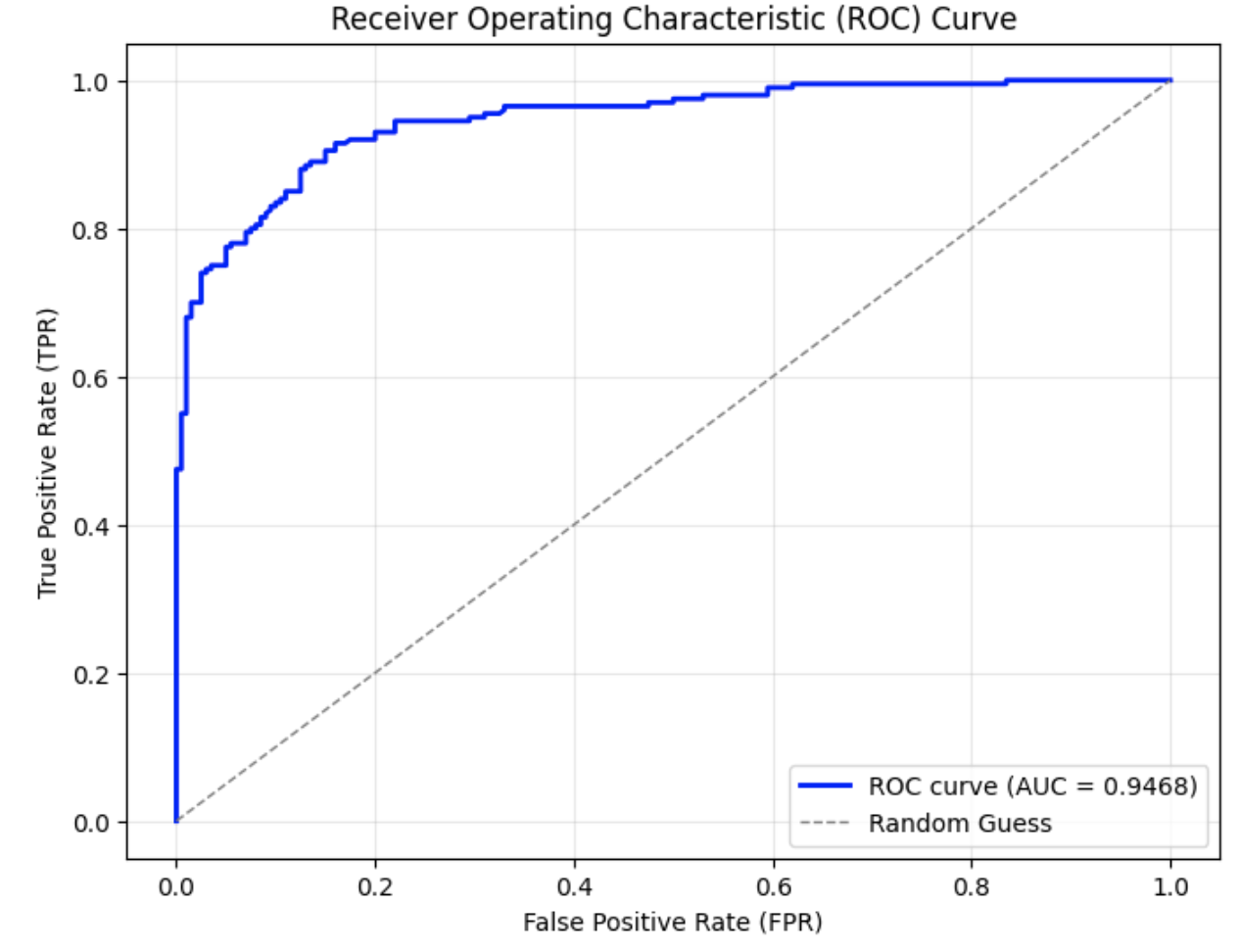}
        \caption{Dipper paraphrasing with $lex\_diversity = 5$}
        \label{fig:dipper_5}
    \end{subfigure}

    \vspace{0.5cm}

    \begin{subfigure}[b]{0.465\linewidth}
        \includegraphics[width=\linewidth]{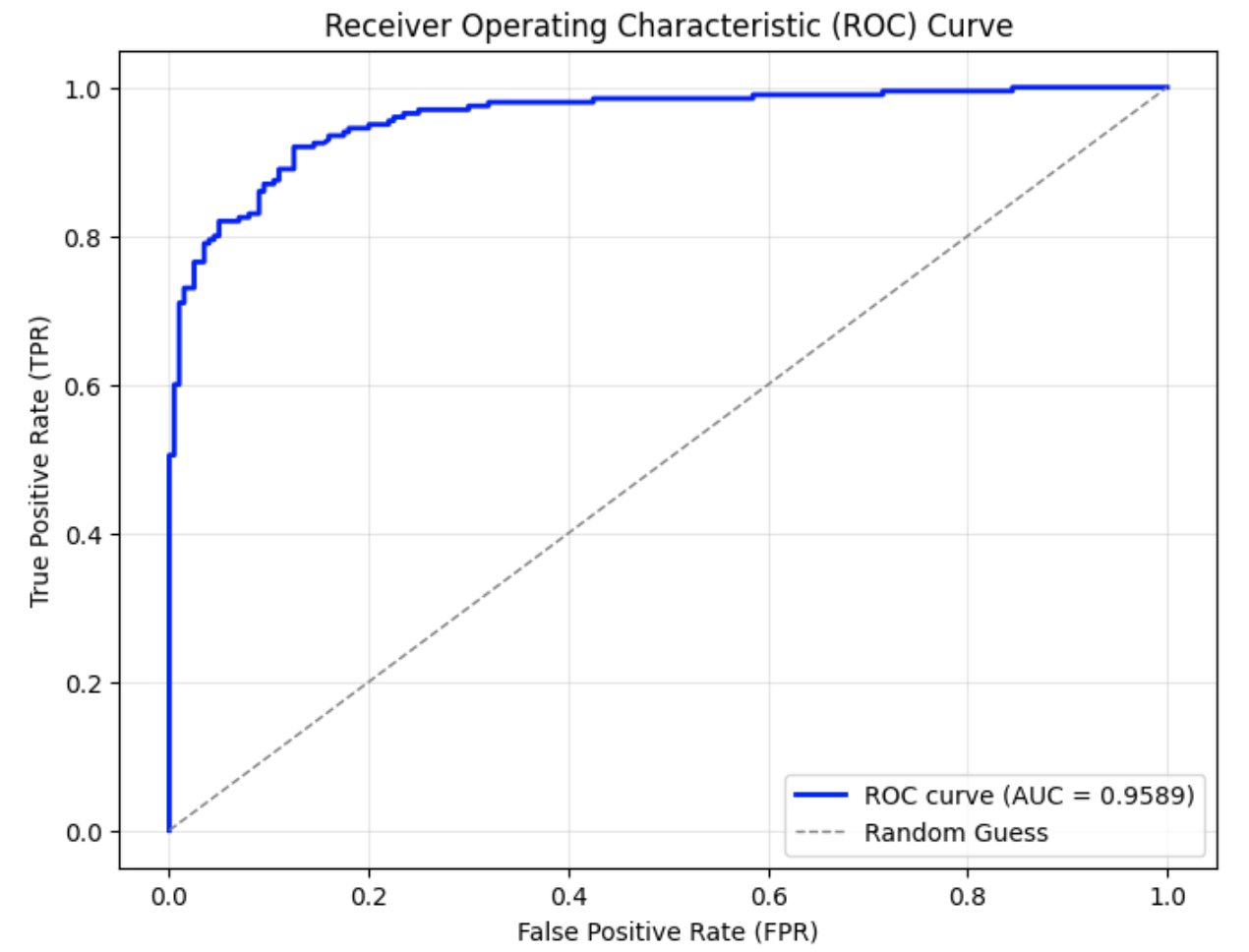}
        \caption{Dipper paraphrasing with $lex\_diversity = 10$\textcolor{white}{aaaaaaa aaaaaaaa}}
        \label{fig:dipper_10}
    \end{subfigure}
    \hfill
    \begin{subfigure}[b]{0.465\linewidth}
        \includegraphics[width=\linewidth]{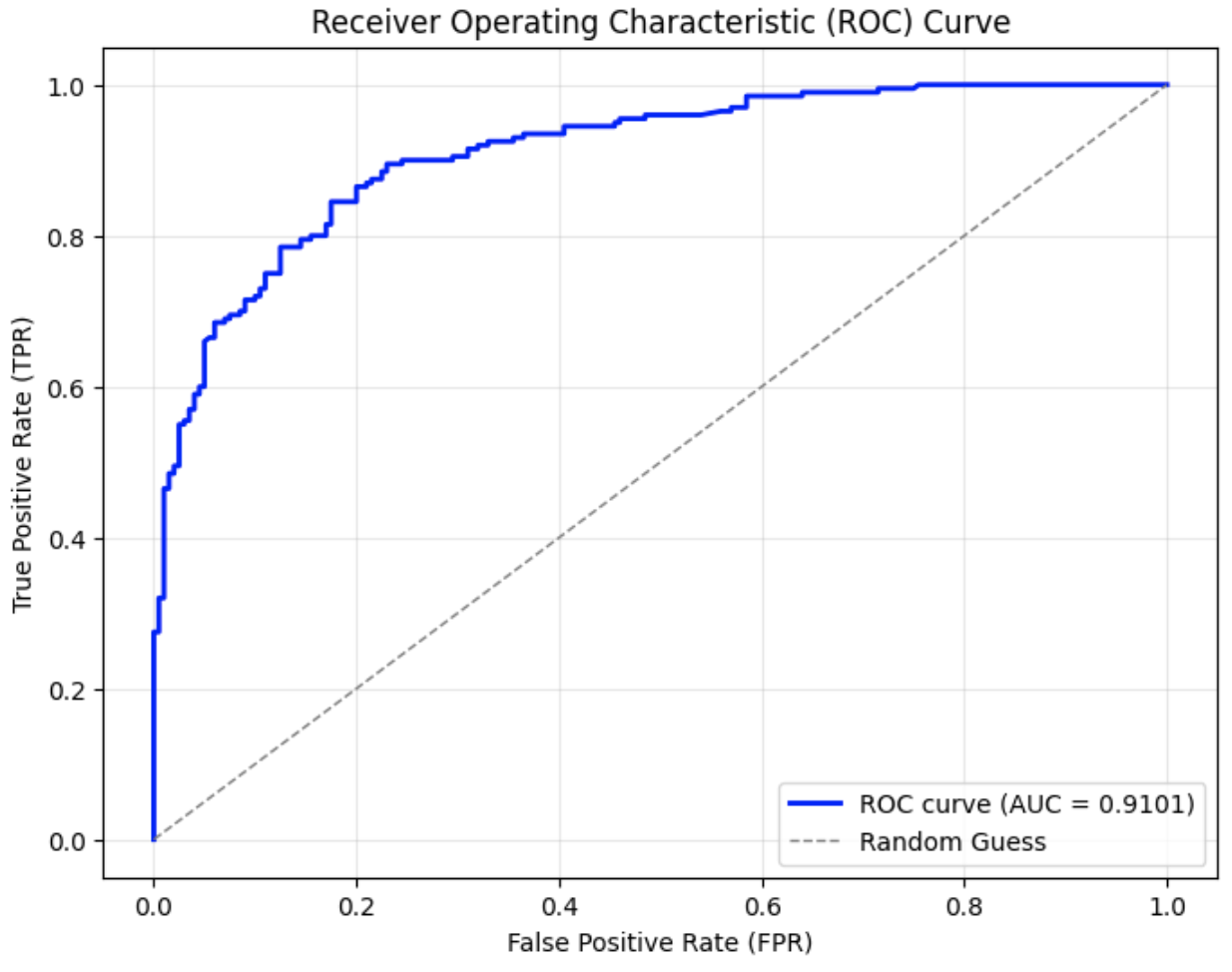}
        \caption{Dipper paraphrasing with $lex\_diversity = 10$ and $order\_diversity = 5$}
        \label{fig:dipper_10_5_synthid}
    \end{subfigure}

    \caption{ROC curves under paraphrasing attacks with different settings.}
    \caption*{\textbf{Note\(^*\):} Due to hardware limitations in Google Colab Pro—specifically, a maximum GPU memory of 40 GB—Dipper could only be run once per session. As a result, the ROC curves were generated in separate runs, requiring a restart between each execution, and are presented across multiple graphs.}
    \label{fig:roc_combined_4}
\end{figure}

In this experiment, we adopted the Dipper paraphrasing model~\cite{krishna2023paraphrasing}, which is built on the T5-XXL~\cite{raffel2020exploring} architecture. Dipper allows fine-tuned control over both lexical and order diversity through configurable parameters. Two levels of lexical diversity were used to conduct the attacks, and the results are shown in Fig.~\ref{fig:roc_combined_4}.

From the graphs, it can be observed that compared to the original ROC curve of SynthID-Text without attack in Fig.~\ref{fig:roc_combined_4}(a), the AUC in Fig.~\ref{fig:roc_combined_4}(b) and (c) decrease by approximately 0.04–0.05 when only lexical diversity was applied. When both lexical diversity and order diversity were set simultaneously, the AUC experienced a decline to 0.91 in Fig.~\ref{fig:roc_combined_4}(d) from 1.00 in the no attack setting.
The corresponding FPR and F1 scores are presented in Table~\ref{tab:paraphrase_attacks_summary_synthID}. Particularly, when \texttt{lex\_diversity=10} and \texttt{order\_diversity=5} (shown in the fourth row), the FPR exceeded 20\%, and the F1 score dropped to 0.84, indicating a significant reduction in detection accuracy under this paraphrasing condition.

\begin{table}[tb]
    \centering
    \caption{Watermark detection accuracy under different paraphrasing attack settings}
    \label{tab:paraphrase_attacks_summary_synthID}
    \begin{tabular}{lccc}     
        \toprule
        Attack    &  TPR   & FPR  &  F1 with best threshold \\
        \midrule
        No attack    &  1.0  &  0.0 &  1.0 \\
        \addlinespace[0.5ex]
        Dipper-5     & 0.915  & 0.16 &  0.882 \\
        \addlinespace[0.5ex]
        Dipper-10    & 0.92   & 0.125 &  0.8998 \\
        \addlinespace[0.5ex]
        Dipper-10-5  & 0.895    & 0.23 &  0.842 \\
        \bottomrule
    \end{tabular}
    \caption*{\textbf{Note\(^*\):} In this figure, \textit{Dipper-$x$} denotes that the Dipper model was run with a lexical diversity parameter of $x$, while \textit{Dipper-$x$-$y$} indicates a lexical diversity of $x$ and an order diversity of $y$.}
\end{table}

\subsection{Re-Translation Attack}
\label{4.5}

\begin{figure}[tb]
    \centering
    \includegraphics[width=0.8\linewidth]{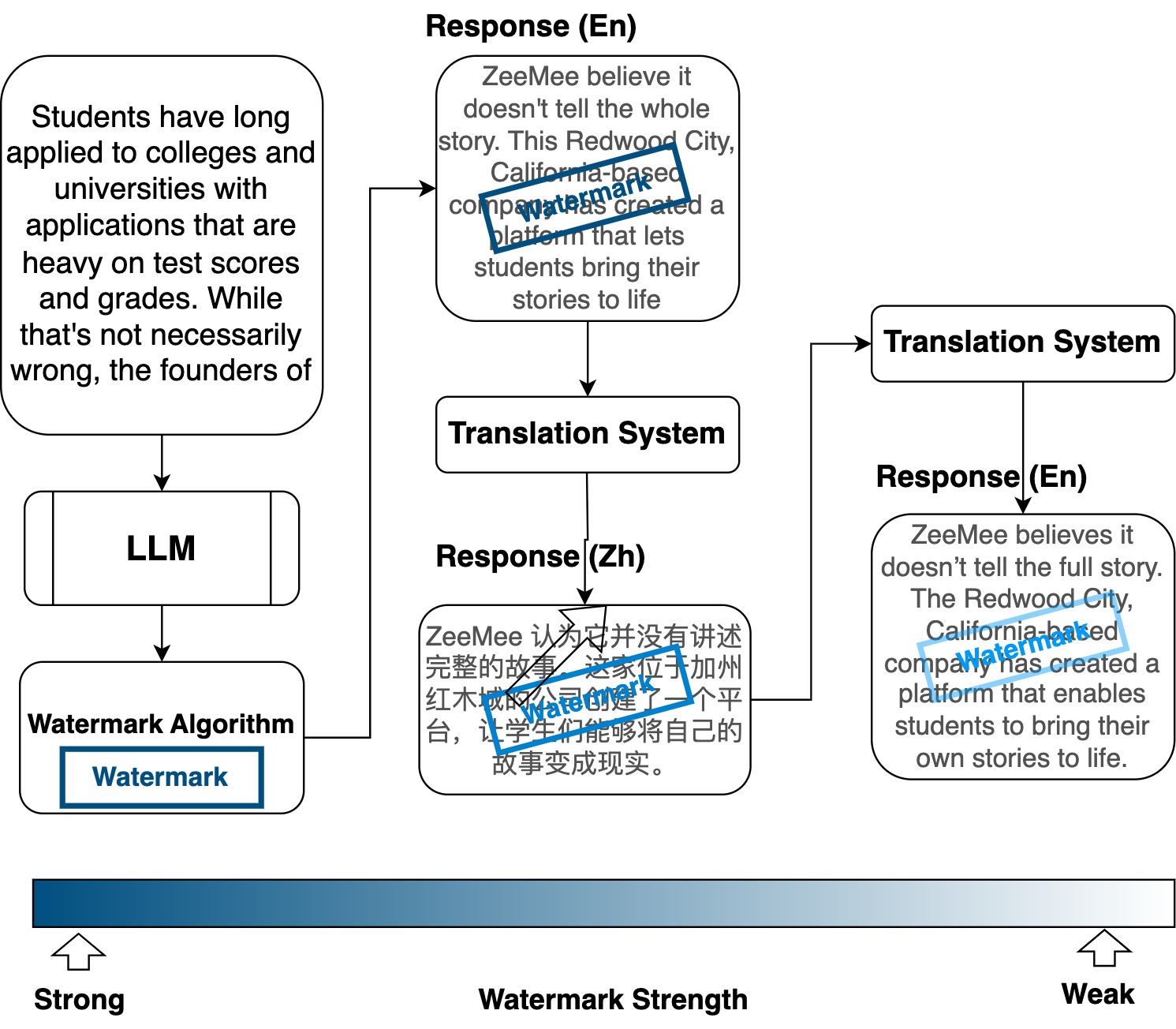}
    % \caption{Watermark dilution through translation}
    \caption{Illustration of watermark dilution through translation}
    \label{fig:watermark_dilution_through_translation}
\end{figure}

\indent The re-translation attack involves translating the original watermarked text into a pivot language and then translating it back into the original language. This process preserves the overall meaning, but may disrupt the watermark signal due to intermediate transformations applied by a translation model, as illustrated in Fig. ~\ref{fig:watermark_dilution_through_translation}.

For this experiment, we used the \texttt{nllb-200-distilled-600M}\footnote{\url{https://huggingface.co/facebook/nllb-200-distilled-600M}} model, a distilled 600M-parameter variant of NLLB-200~\cite{costa2022no}. NLLB-200 is a multilingual machine translation model that supports direct translation between 200 languages and is designed for research purposes. Several different languages were selected as pivot languages, including French, Italian, Chinese, and Japanese. Since the original dataset only consists of English prompts and human-written English completions, the watermarked outputs were first translated into pivot language and then re-translated into English to maintain consistency with the original prompt language.

\begin{figure}
    \centering
    \includegraphics[width=0.8\linewidth]{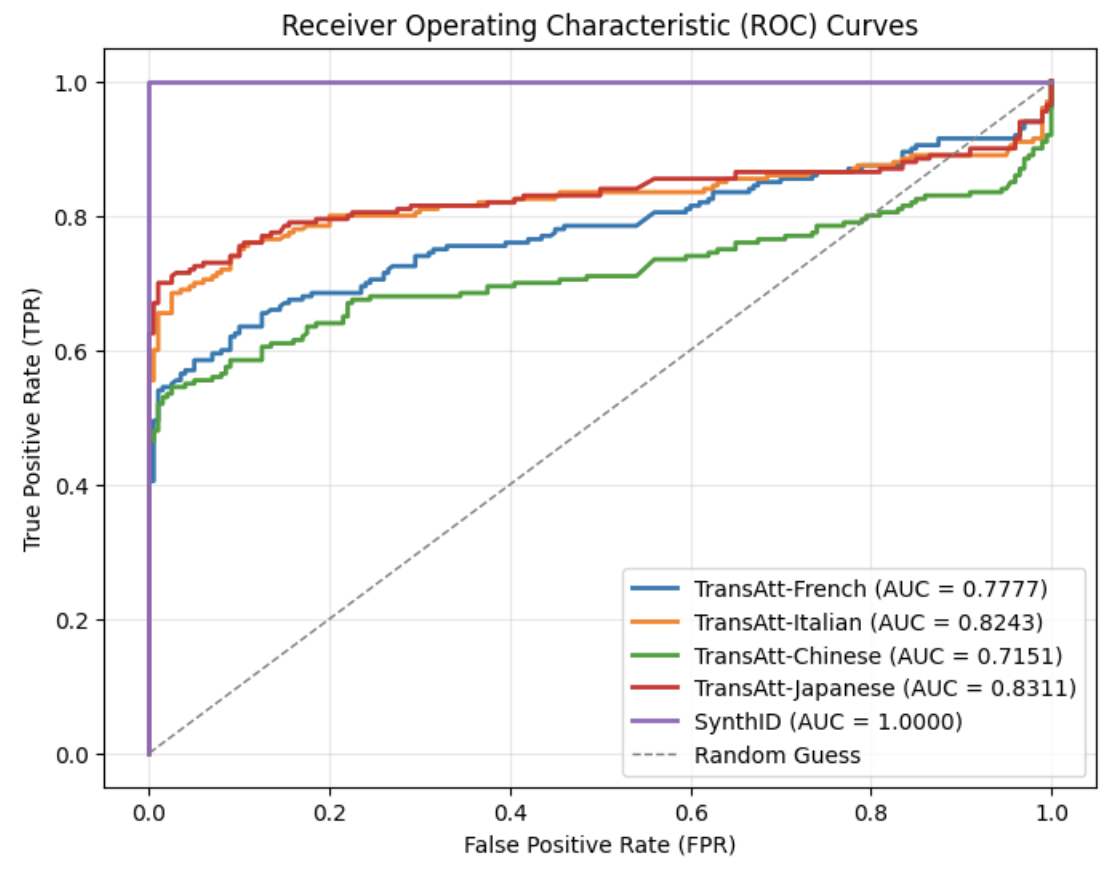}
    \caption{ROC curves of re-translation attacks on SynthID}
    \label{fig:Retranslation attacks for SynthID-Text}
\end{figure}

The ROC curves under this re-translation attacks using different pivot languages are presented in Fig. \ref{fig:Retranslation attacks for SynthID-Text}. 
The results indicate that the choice of pivot language significantly influences the effectiveness of re-translation attacks. French and Italian, which both belong to the Latin language family , share substantial linguistic similarities with English, which has been heavily influenced by Latin. As a result, the round-trip retranslated texts maintain relatively high AUC scores. In contrast, Chinese is more significantly different from English, leading to the lowest AUC observed after re-translation. Surprisingly, Japanese produces the highest AUC among all tested pivot languages, even slightly surpassing Italian. This outcome may be attributed to the specific design of English-to-Japanese translation systems. Given the syntactic differences between Japanese and English (such as SOV versus SVO word order), many modern translation tools adopt a linear translation strategy when translating from English to Japanese\cite{mizowaki2022syntactic, sekizawa2017improving}. This approach attempts to preserve the original sentence structure as much as possible to enhance translation quality. Consequently, round-trip translation using Japanese tends to retain more of the original semantics and structure, making the re-translation attack less effective. 
Compared to the baseline performance of SynthID-Text without attack, the F1 score for the re-translation attack using Chinese reduces significantly from 1.00 to 0.711, while the F1 score remains 0.819 for Japanese, which is the highest, as shown in Table~\ref{tab:Retranslation_summary_synthID}. 
% The FPR increases from 0.00 to 0.035, and the F1 score declines noticeably from 1.00 to 0.71, as shown in Table~\ref{tab:Retranslation_summary_synthID}, indicating a substantial degradation in watermark detectability under this attack scenario.

% \begin{figure}[htb!]
%     \centering
%     \includegraphics[width=0.7\linewidth]{figures/re-translation attack roc curve.png}
%     % \caption{ROC curve under re-translation attack using nllb-200-distilled-600M model.}
%     \caption{ROC curve under re-translation attack using the nllb-200-distilled-600M model}
%     \label{fig:re-translation roc}
% \end{figure}

% \begin{table}[tb]
%     \centering
%     \caption{Watermark detection accuracy under re-translation attack}
%     \label{tab:Retranslation_summary_synthID}
%     \begin{tabular}{lllc}     
%         \toprule
%         Attack               &  TPR   & FPR  &  F1 with best threshold \\
%         \midrule
%         No attack         & 1.0   & 0.0 &  1.0 \\
%         Retranslation     & 0.575  & 0.035 &  0.714 \\
%         \bottomrule
%     \end{tabular}
% \end{table}
\begin{table}[tb]
    \centering
    \caption{Watermark detection accuracy under re-translation attacks using different pivot languages}
    \label{tab:Retranslation_summary_synthID}
    \begin{tabular}{lccc}     
        \toprule
        Attack        &  TPR   & FPR  &  F1 \\
        \midrule
        No attack         & 1.0   & 0.0 &  1.0 \\
        Re-trans-French     & 0.675  & 0.155 &  0.738 \\
        Re-trans-Italian     & 0.76  & 0.11 &  0.813 \\
        Re-trans-Chinese     & 0.675  & 0.225 &  0.711 \\
        Re-trans-Japanese    & 0.715  & 0.03 &  0.819 \\
        \bottomrule
    \end{tabular}
\end{table}

\subsection{Summary}
\label{4.6}

Table~\ref{tab:attack_summary} summarizes the watermark detection performance of SynthID-Text under various attack scenarios. For the re-translation attack, we present the result for Chinese as it is one of the three most widely spoken languages in the world. 

Without any attack, the algorithm achieves a perfect F1 score of 1.0 and a false positive rate (FPR) of 0.0, demonstrating excellent baseline performance in detecting watermarked text.
Under synonym substitution attacks, the F1 score decreases to 0.884, slightly below 0.9, indicating a moderate level of resilience to lexical variation. 

For the copy-and-paste attack with a length ratio of 10, the F1 score decreases more substantially to 0.788, while the FPR rises sharply to 0.53. This suggests that simply appending large segments of natural (unwatermarked) text can significantly weaken watermark detectability, even if the original watermarked content remains unchanged.
The paraphrasing attack, particularly when involving both high lexical diversity (\texttt{lex\_diversity} = 10) and syntactic reordering (\texttt{order\_diversity} = 5), also lead to a notable decrease in robustness. In this setting, the FPR increases to 0.23, and the F1 score falls to 0.842.

The most severe degradation occurs under the re-translation attack. Translating the watermarked text into Chinese and subsequently back into English results in a significant decline in detection performance: the F1 score falls to 0.711, and the TPR declines to 0.675, only slightly better than random guessing. This highlights the substantial vulnerability of SynthID-Text to semantic-preserving transformations.

\textbf{These findings suggest that while SynthID-Text remains robust against simple lexical substitutions, it is significantly less effective under complex semantic-preserving attacks such as paraphrasing and round-trip translation, which pose the greatest challenges for reliable watermark detection.}

\begin{table}[tb]
    \centering
    \caption{Watermark detection accuracy of SynthID-Text under various attacks}
    \label{tab:attack_summary}
    \begin{tabular} {lccc}
    % {>{\centering\arraybackslash}m{5.5cm} 
    %             >{\centering\arraybackslash}m{2cm} 
    %             >{\centering\arraybackslash}m{2cm} 
    %             >{\centering\arraybackslash}m{3cm}}
        \toprule
        Attack & TPR & FPR & F1 \\
        \midrule
        No attack & 1.0 & 0.0 & 1.0 \\
        \addlinespace[0.7ex]
        Substitution ($\epsilon = 0.7$) & 0.82 & 0.035 & 0.884 \\
        \addlinespace[0.7ex]
        Copy-and-Paste (ratio = 10) & 0.995 & \textcolor{red}{0.53} & 0.788 \\
        \addlinespace[0.8ex]
        Paraphrasing (\texttt{lex\_diversity} \\ = 10, \texttt{order\_diversity} = 5) & 0.895 & 0.23 & 0.842 \\
        \addlinespace[0.85ex]
        Re-Translation (Chinese) & \textcolor{red}{0.675} & 0.225 & \textcolor{red}{0.711} \\
        \bottomrule
    \end{tabular}
\end{table}

\section{SynGuard: An Enhanced SynthID-Text Watermarking}
\label{ch:evaluation}

Since SynthID-Text embeds watermarks during the text generation process, if the generated text is regenerated or modified by another translation or language model, the original watermarking signals may be disrupted. As a result, the watermark information is prone to being destroyed. This vulnerability becomes especially apparent in the detection performance when subjected to back-translation attacks. The results could be found in Section \ref{ch:results for SynGuard}.%, as shown in Table~\ref{tab:attack_summary}.

In this section, we introduce a novel watermarking method, SynGuard, which combines the Semantic Invariant Robust (SIR) watermarking algorithm \cite{liu2024semanticinvariantrobustwatermark} with the SynthID-Text tournament sampling mechanism \cite{dathathri2024scalable}. %Subsection~\ref{5.1} provides an overview of the proposed method. Subsection~\ref{5.2} details the detection process for SynGuard. 

% Finally, Section~\ref{5.3} presents an ablation study to analyze the contribution of model parameter.

% 参考morphmark的方案建模，tabluarmark的安全性分析
% Fig. 7只提到了水印嵌入（该图暂时保留，用以再水印嵌入时说明用）。overview应该包括水印检测部分(左边水印嵌入，中间攻击信道，右边水印提取)

\subsection{Watermark Embedding}
\label{5.1}

Watermarking algorithms embed watermarks by modifying logits during the token generation process. SynthID-Text achieves this by using the hash values of preceding tokens along with a secret key $k$ to generate pseudorandom numbers. 
% These numbers are then used to 
% % partition the vocabulary into "green" and "red" lists. 
% impact the token sampling process.
These numbers are then used to guide the token sampling process.
This design, based on pseudorandom functions and a fixed key, makes the watermark difficult to remove unless the attacker has access to both the key and the random seed.

However, if the entire text is regenerated by another language model, such as in the back-translation scenario, the watermark signal can be severely degraded. This vulnerability stems from the fact that SynthID-Text does not incorporate semantic understanding into its watermarking process. By contrast, the SIR algorithm~\cite{liu2024semanticinvariantrobustwatermark} embeds watermark signals by mapping semantic features of preceding tokens to specific token preferences. This semantic-aware approach has demonstrated resilience to meaning-preserving transformations.

%下面的内容与上面部分稍显重复，需要突出引入了新的logits偏置，去除多余内容。目前核心在于sir的logits是使用的神经网络，而xia han目前没有在论文中提 (核心缺陷)
To enhance robustness against semantic perturbations, we propose a hybrid approach that integrates SynthID-Text with SIR. This new method, called \textbf{SynGuard}, generates three separate sets of logits at different stages and combines them to form the final logits vector. This vector is then passed through a softmax function to obtain a probability distribution over the vocabulary \(V\). The three component logits are:
\begin{itemize}
    \item \textbf{Base LLM logits:} Generated directly from the backbone LLM, representing the standard token probabilities.
    \item \textbf{SIR logits:} Derived from a semantic watermarking model conditioned on the preceding text, encoding semantic consistency.
    \item \textbf{SynthID logits:} Computed using the pseudorandom watermarking mechanism based on hash values of tokens, a random seed and a secret key.
\end{itemize}

\begin{figure}[tb]
    \centering
    \includegraphics[width=1\linewidth]{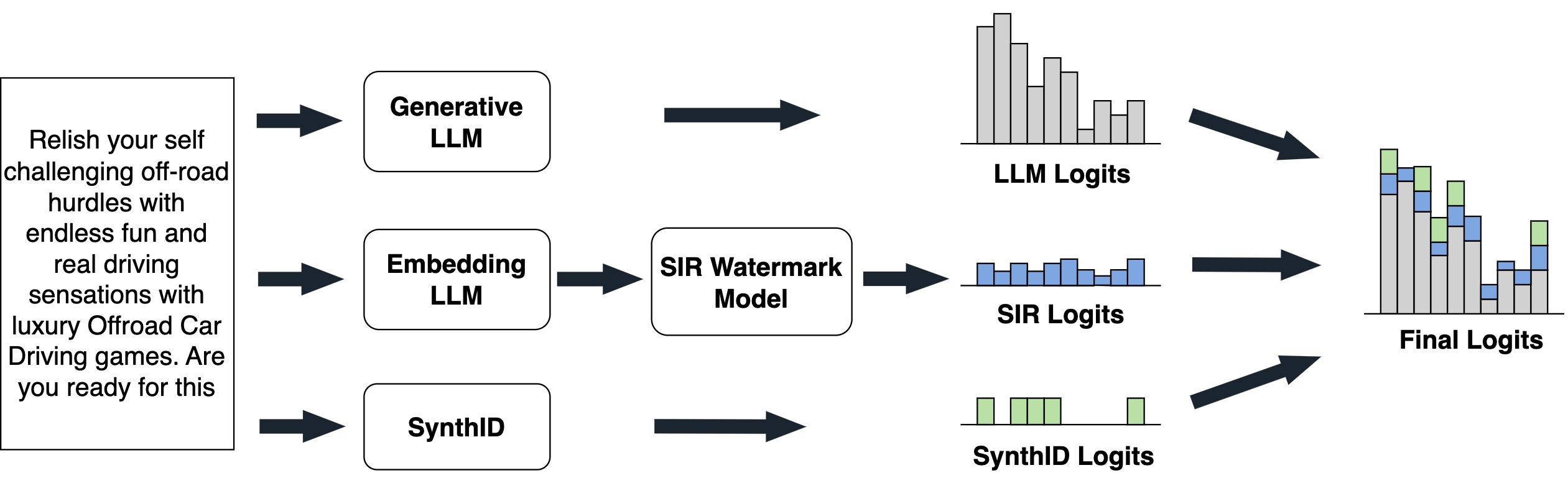}
    \caption{SynGuard watermark embedding.}
    \label{fig:new_algorithem_overview}
    %\vspace{-0.15in}
\end{figure}

The overall embedding process is illustrated in Fig.~\ref{fig:new_algorithem_overview}, and the detailed procedure is described in Algorithm~\ref{algo:watermark}.

% \begin{algorithm}
%     \caption{Watermark Generation}
%     \label{algo:algo_1}
%     \begin{algorithmic}[1]
%         \Require{a language model M, a prompt $x^{prompt}$, previous generated text $ t = [t_0, t_1,... ,t_{T-1}]$, a text embedding model $E$, trained watermark model $ W $, a semantic weight $ \delta $, a tournament sampling algorithm $ G $, a key $k$, a token $x$ from vocabulary.}
%         \Statex
%         \State Generate next token logits from language model M, get $P_M(x^{prompt}, t_{:T-1})$
%         \State Generate the embedding of the text $e_{:T-1}$: $E_{:T-1}$
%         \State Generate SIR watermark logits from trained watermark model $W$: $P_W(E_{:T-1})$
%         \State Generate SynthID-Text watermark logits through tournament sampling function $G$:$P_G(x^{prompt}, k, x)$
%         \State Get the final logits $P_{\hat{M}}$ with the given text $ t $:
%         \begin{align*}
%             P_{\hat{M}}(x^{prompt}, t_{:T-1}) = P_M(x^{prompt}, t_{:T-1}) + \delta * P_W(E_{:T-1}) + (1-\delta) * P_G(x^{prompt}, k, x)
%         \end{align*}
%         \Statex
%         \Ensure{watermarked next token logits $P_{\hat{M}}$($t_{T}$)}
%     \end{algorithmic}
% \end{algorithm}
\begin{algorithm}
    \caption{Watermark Embedding of SynGuard}
    \label{algo:watermark}
    \begin{algorithmic}[1]
        \Require Language model $M$, prompt $x^{\text{prompt}}$, text $t = [t_0, ..., t_{T-1}]$, embedding model $E$, watermark model $W$, semantic weight $\delta$, tournament sampler $G$, key $k$, token $x$
        % \Statex
        \State Generate logits from $M$: $P_M(x^{\text{prompt}}, t_{:T-1})$;
        \State Generate embedding $E_{:T-1}$;
        \State Get SIR watermark logits $P_W(E_{:T-1})$;
        \State Get SynthID-Text watermark logits $P_G(x^{\text{prompt}}, k, x)$;
        \State Compute:
        \begin{align*}
        P_{\hat{M}}(x^{\text{prompt}}, t_{:T-1}) &= P_M(x^{\text{prompt}}, t_{:T-1}) \\
        &\quad + \delta \cdot P_W(E_{:T-1}) \\
        &\quad + (1 - \delta) \cdot P_G(x^{\text{prompt}}, k, x).
        \end{align*}
        % \Statex
        \Ensure Final watermarked logits $P_{\hat{M}}(t_T)$
    \end{algorithmic}
\end{algorithm}
%换言之，可以归结为双水印

\subsection{Watermark Extraction}
\label{5.2}

SynGuard determines whether a given text is watermarked by evaluating both the semantic similarity to the preceding context and the statistical watermark signal encoded as $g$-values. 
Intuitively, the more semantically aligned a token is with its context, and the higher its corresponding $g$-value, the more probable it is that the text was generated by a watermarking algorithm.

\textbf{Watermark Strength.} The probability that a text contains a watermark is quantified by a composite score \textit{s}. A higher \textit{s} indicates a higher probability that the text is watermarked. Given a text $t = [t_0, t_1, \ldots, t_T]$, we compute two components:

\begin{itemize}
    \item \textbf{Semantic similarity score:} Let $P_W(x_i, t_{:T-1})$ denote the semantic similarity between the token and the preceding generated text, computed using a pretrained semantic watermark model $W$. The normalized semantic score is:
    \[
    s_{\text{semantic}} = \frac{1}{T} \sum\limits{}_{i=0}^{T} \left(P_W(x_i, t_{:T-1}) - 0 \right).
    \]
    \item \textbf{G-value score:} Let $g_l$ represent the output of the \(l_{th}\)  SynthID-Text watermarking function for tokens. The average $g$-value score is:
    \[
    s_{\text{g-value}} = \frac{1}{T*m} \sum_{i=0}^{T} \sum_{l=0}^{m} g_l (x_i,t_{:T-1}).
    \]
\end{itemize}

Since $s_{\text{semantic}} \in [-1, 1]$ and $s_{\text{g-value}} \in [0, 1]$, we normalize $s_{\text{semantic}}$ to fall within the same range by applying a linear transformation. The final score \(s\) is computed as:

\begin{equation}
\label{eq:z_score}
s = \delta \cdot \frac{s_{\text{semantic}} + 1}{2} + (1 - \delta) \cdot s_{\text{g-value}}.
\end{equation}

Here, $\delta \in [0,1]$ is a hyperparameter that controls the relative weighting between the semantic similarity signal and the token-level watermark signal. A larger $\delta$ places more emphasis on semantic alignment, while a smaller $\delta$ favors the token sampling randomness.

% TODO 这个小标题需要删除 ？？
\subsection{Robustness Analysis}
\label{sec:robustness_analysis}
To evaluate the robustness of \textbf{SynGuard}, we consider adversaries who attempt to remove or forge the watermark while preserving the underlying semantics. Our hybrid approach combines semantic-awareness from SIR and pseudorandom unpredictability from SynthID, offering both attack robustness and key-based security guarantees.
%定理1. 我们的方案对于单词删除、替换类攻击很鲁棒。
%证明：我们的方案偏置主要包括两部分：P_{\hat{M}}(x^{\text{prompt}}, t_{:T-1}) = P_M(x^{\text{prompt}}, t_{:T-1})  \quad + \delta \cdot P_W(E_{:T-1}) \quad + (1 - \delta) \cdot P_G(x^{\text{prompt}}, k, x).
%普通单词替换类攻击会显著影响P_G(x^{\text{prompt}}, k, x)部分
\begin{theorem}
Let $t = [t_0, t_1, \ldots, t_T]$ be a watermarked text and $t'$ be a meaning-preserving transformation of $t$. Then, with high probability, the watermark detection score $s(t')$ remains above detection threshold $\tau$, i.e., the watermark is still detectable.
\end{theorem}

\begin{proof}
%简易版证明：
% The detection score $s$ is defined as: $
% s = \delta \cdot \frac{s_{\text{semantic}} + 1}{2} + (1 - \delta) \cdot s_{\text{g-value}}.$ Under meaning-preserving transformations:
% \begin{itemize}
%     \item \textbf{Semantic component:} Since $t'$ preserves meaning, the semantic embedding $E(t_{:i})$ and $E(t'_{:i})$ are close. Assuming the semantic watermark model $W$ is Lipschitz continuous:
%     \[
%     |P_W(E(t_{:i})) - P_W(E(t'_{:i}))| \le L \cdot \|E(t_{:i}) - E(t'_{:i})\|.
%     \]
%     Thus, $s'_{\text{semantic}} \approx s_{\text{semantic}}$.
    
%     \item \textbf{g-value component:} Without access to key $k$, the attacker cannot reproduce SynthID preferences. Hence, $s'_{\text{g-value}} \approx 0.5$.
% \end{itemize}

% As long as $\delta$ is reasonably large, $s(t')$ stays above $\tau$ because $s_{\text{semantic}}$ dominates and remains high. Hence, the watermark survives semantic-preserving attacks.
% 扩展版：
The detection score $s$ is a weighted sum of two components: a semantic alignment score $s_{\text{semantic}}$ and a pseudorandom signature score $s_{\text{g-value}}$.

%\begin{itemize}
%\textbf{Semantic Alignment:} 
Because $t'$ preserves the meaning of $t$, the contextual embeddings of $t'$ remain close to those of $t$. Let $E(t_{:i})$ denote the semantic embedding of the prefix up to token $t_i$. Since $t'$ has nearly the same context at each position in a semantic sense, we have $\|E(t_{:i}) - E(t'_{:i})\|$ small for all $i$. The semantic watermark model $W$ is assumed to be Lipschitz continuous \cite{liu2024semanticinvariantrobustwatermark}: 
$$|P_W(E(t_{:i})) - P_W(E(t'_{:i}))| \le L \cdot \|E(t_{:i}) - E(t'_{:i})\|,$$
where $L>0$ denotes the Lipschitz constant. 

In other words, the watermark bias for the next token does not drastically change under a semantically invariant perturbation. Consequently, for each token position $i$, the semantic preference $P_W(x_i, t_{:i-1})$ assigned by $W$ to the actual token $x_i$ in $t'$ will be close to the value it was for $t$. If $t$ was watermarked, most tokens had high semantic preference values (the watermark favored those choices); $t'$, using synonymous or rephrased tokens, will on average still yield high $P_W$ values for each token, since the tokens remain well-aligned with a similar context. Thus, for each token $x'_i$ in $t'$, we get
\[
s'_{\text{semantic}} = \frac{1}{T} \sum\limits{}_{i=0}^{T} \left(P_W(x'_i, t'_{:i-1}) - 0\right) \approx s_{\text{semantic}} - \varepsilon,
\]
for some small $\varepsilon$.

%\textbf{Pseudorandom Signature:} 
The SynthID component uses a secret key $k$ to generate pseudorandom preferences. Without $k$, $s'_{\text{g-value}} \approx 0.5$. In the original watermarked $t$, tokens are biased toward higher $g$-values. Hence, under semantic-preserving transformation, the g-value component drops to 0.5, but $s_{\text{semantic}}$ remains high.
%\end{itemize}

Therefore, the overall score:$s(t') = \delta \cdot \frac{s'_{\text{semantic}} + 1}{2} + (1 - \delta) \cdot s'_{\text{g-value}}$ is still above threshold if $\delta$ is reasonably large. In conclusion, the watermark remains detectable in $t'$.
\end{proof}

\begin{theorem}
Let $k$ be the watermark key for SynGuard. For any text $u$ not generated by the watermarking algorithm, the probability that $s(u) > \tau$ is exponentially small in $T$.
\end{theorem}

\begin{proof}
%简易版证明
% \begin{itemize}
%     \item \textbf{Unforgeability:} For $T$ tokens, the probability that random token choices align with the g-value 1 is bounded by:
%     \[
%     \Pr\left(\frac{1}{T} \sum_{i=1}^T G_i \ge p_{\text{wm}} \right) \le \exp\left(-2T(p_{\text{wm}} - 0.5)^2\right).
%     \]
%     \item \textbf{False Positive Rate:} Let $X_i, Y_i$ denote $g$-value and semantic score respectively. For unwatermarked text:
%     \[
%     \mathbb{E}[s(u)] = 0.5,\quad \text{Var}[s(u)] = O\left(\frac{1}{T}\right).
%     \]
%     Using Chebyshev’s inequality:
%     \[
%     \Pr(|s(u) - 0.5| > \eta) \le \frac{\text{Var}[s(u)]}{\eta^2} = O\left(\frac{1}{T\eta^2}\right).
%     \]
%     Therefore, $\Pr(s(u) > \tau)$ becomes negligible.
% \end{itemize}
The robustness stems from the pseudorandom behavior of the SynthID component, which introduces a hidden bias into token selection based on a watermark key $k$. The watermarking model adds a preference signal $g_k(x_i, t_{:T-1}) \in [0,1]$ for candidate tokens, and combines it with the semantic alignment score $P_W$. The detector computes a combined score:
\begin{align*}
    s =  \frac{\delta}{T} \sum_{i=1}^T  \frac{P_W(x_i, t_{:T-1})+1}{2} +  \frac{(1-\delta)}{T} \sum_{i=1}^T g_k(x_i, t_{:T-1}).
    % s =  \frac{\delta}{T} \sum_{i=1}^T \left( \frac{P_W(x_i, t_{:T-1})+1}{2} \right) +  \frac{(1-\delta)}{T} \sum_{i=1}^T g_k(x_i, t_{:T-1}).
\end{align*}

Now consider an attacker attempting to generate a fake watermarked text without access to $k$:
\begin{itemize}
    \item Since $g_k$ is keyed and pseudorandom, its outputs are statistically independent of the attacker’s choices.
    \item Therefore, the second term in $s$, the SynthID component, behaves like uniform noise with expected value $\approx 0.5$ and variance $O(1/T)$.
    \item The first term (semantic preference) is not optimized in the attacker’s text either, since only the original watermarker uses $P_W$ for guidance.
    \item Hence, the attacker’s overall score $s_{\text{fake}} \approx 0.5$, with small deviations bounded by concentration inequalities.
\end{itemize}

Let $Y_i = \frac{P_W(x_i, t_{:i-1}) + 1}{2}$ and $Z_i = g_k(x_i, t_{:i-1})$, both taking values in $[0,1]$. Define $X_i := \delta Y_i + (1 - \delta) Z_i$, so $X_i \in [0,1]$. Since $g_k$ is pseudorandom with no attacker control, and $P_W$ is optimized only during watermark generation, their expected values over attacker-generated text are both approximately $0.5$. Hence $\mathbb{E}[X_i] = 0.5$. With $\mathbb{E}[X_i] = 0.5$, and $X_1, \dots, X_T$ are i.i.d., Hoeffding’s inequality gives:
% Let $Y_i = \frac{P_W(x_i, t_{:i-1}) + 1}{2}$ and $Z_i = g_k(x_i, t_{:i-1})$, both bounded in $[0,1]$. Define $X_i := \delta Y_i + (1 - \delta) Z_i$, so $X_i \in [0,1]$. Assuming the $X_i$ are i.i.d. with $\mathbb{E}[X_i] = 0.5$, Hoeffding’s inequality gives:
$$
\Pr(s > \tau) = \Pr\left( \frac{1}{T} \sum_{i=1}^T X_i > \tau \right)
\le e^ { -2T(\tau - 0.5)^2 }.
$$

This shows that for any non-watermarked text $u$, the probability of it being misclassified as watermarked (i.e., $s(u) > \tau$) decays exponentially with length $T$.

% Formally, let each score term be modeled as i.i.d. variables $X_i \sim \text{Unbiased}$, then:
% \[
% \mathbb{E}[s_{\text{fake}}] = 0.5, \quad
% \text{Var}(s_{\text{fake}}) = O\left(\frac{1}{T}\right).
% \]

% Thus, for any margin $\epsilon > 0$, we have:
% \[
% \Pr(|s_{\text{fake}} - 0.5| > \epsilon) \le \frac{C}{T \epsilon^2}.
% \]

Meanwhile, a genuine watermarked text has both components biased upward (semantic tokens aligned and token scores chosen with positive $g_k$ bias), yielding $s_{\text{true}} > \tau$, where $\tau \in (0.6, 0.9)$ is the detection threshold.

Therefore, false positives (attacker's text exceeding threshold) are exponentially rare as $T$ increases. Likewise, removal attempts (via editing tokens) cannot reduce the score below threshold unless semantic meaning is also damaged.
\end{proof}

\section{Experimental Evaluation}
\label{ch:results for SynGuard}
%\label{ch:synthID_results}
%缩减下面内容
This section presents the experimental settings, evaluation metrics, and results of SynGuard compared to the baselines. %Section~\ref{4.1} outlines the experimental setup, including the backbone model, dataset, and metrics used for evaluation. Sections~\ref{4.2} through~\ref{4.5} report our scheme's performance under four types of text editing attacks: synonym substitution, copy-and-paste, paraphrasing, and re-translation. At the same time, we also compare with the baselines in these subsections. %Finally, Section~\ref{4.6} summarizes and compares results across all attack types to provide a comprehensive evaluation.

%We applied the same experimental settings as described in Session~\ref{4.1} to conduct experiments for SynGuard in this chapter. Session~\ref{6.1} presents the detection performance of the proposed SynGuard watermarking algorithm in comparison to SynthID-Text and SIR. ROC-AUC is used to evaluate detection accuracy, while iteration time is reported to assess computational efficiency. The quality of the generated text is assessed using Perplexity (PPL) scores, which reflect how fluent and natural the output text is. Session~\ref{6.2} provides the robustness evaluation results of SynGuard under four types of attacks, synonym substitution, copy-and-paste, paraphrasing, and re-translation, and compares its performance with that of SynthID-Text.

\subsection{Experimental Setup}
\label{4.1}

\indent \textbf{Backbone Model and Dataset.} All experiments were conducted using \texttt{Sheared-LLaMA-1.3B} \cite{xia2023sheared}, a model further pre-trained from \texttt{meta-llama/Llama-2-7b-hf}\footnote{\url{https://huggingface.co/meta-llama/Llama-2-7b-hf}} and \texttt{opt-1.3B}\footnote{\url{https://huggingface.co/facebook/opt-1.3b}} from Meta. These models used are publicly available via HuggingFace.
For the dataset, we adopt the Colossal Clean Crawled Corpus (C4)~\cite{raffel2020exploring}, which includes diverse, high-quality web text. Each C4 sample is split into two segments: the first segment serves as the prompt for generation, while the second (human-written) segment is used as reference text. The quality of the generated text is assessed using Perplexity (PPL) scores, which reflect how fluent and natural the output text is. These unaltered human texts are treated as control data for evaluating the watermark detector’s false positive rate.

\textbf{Evaluation Metrics.} The robustness is evaluated using the following metrics: True Positive Rate (TPR), False Positive Rate (FPR), F1 Score, and ROC-AUC.
%\begin{itemize}
    % \item \textbf{True Positive Rate (TPR):} The proportion of watermarked texts correctly identified.
    % \item \textbf{False Positive Rate (FPR):} The proportion of unwatermarked texts incorrectly identified as watermarked.
    % \item \textbf{F1 Score:} The harmonic mean of precision and recall, computed at the best threshold.
    % \item \textbf{ROC-AUC:} The area under the Receiver Operating Characteristic (ROC) curve, measuring overall classification performance across all thresholds.
%\end{itemize}
Each experiment was conducted using 200 watermarked and 200 unwatermarked samples, each with a fixed length of \( T = 200 \) tokens, as same as the default setting of \cite{pan2024markllm, wang-etal-2025-morphmark}. All experiments were implemented using the MarkLLM toolkit~\cite{pan2024markllm}.

\subsection{Main Results}
\label{6.1}
% \indent This section uses the F1 score to demonstrate the detection accuracy of SynGuard, alongside comparisons with the baseline methods, SIR and SynthID-Text. 
This section uses the F1 score to demonstrate the detection accuracy of SynGuard, and compares it to the baseline methods, SIR and SynthID-Text.
The naturalness of the output texts generated by these three algorithms is also evaluated to assess their textual quality.

\begin{figure}[tb]
    \centering
    \begin{subfigure}[b]{0.47\linewidth}
        \includegraphics[width=\linewidth]{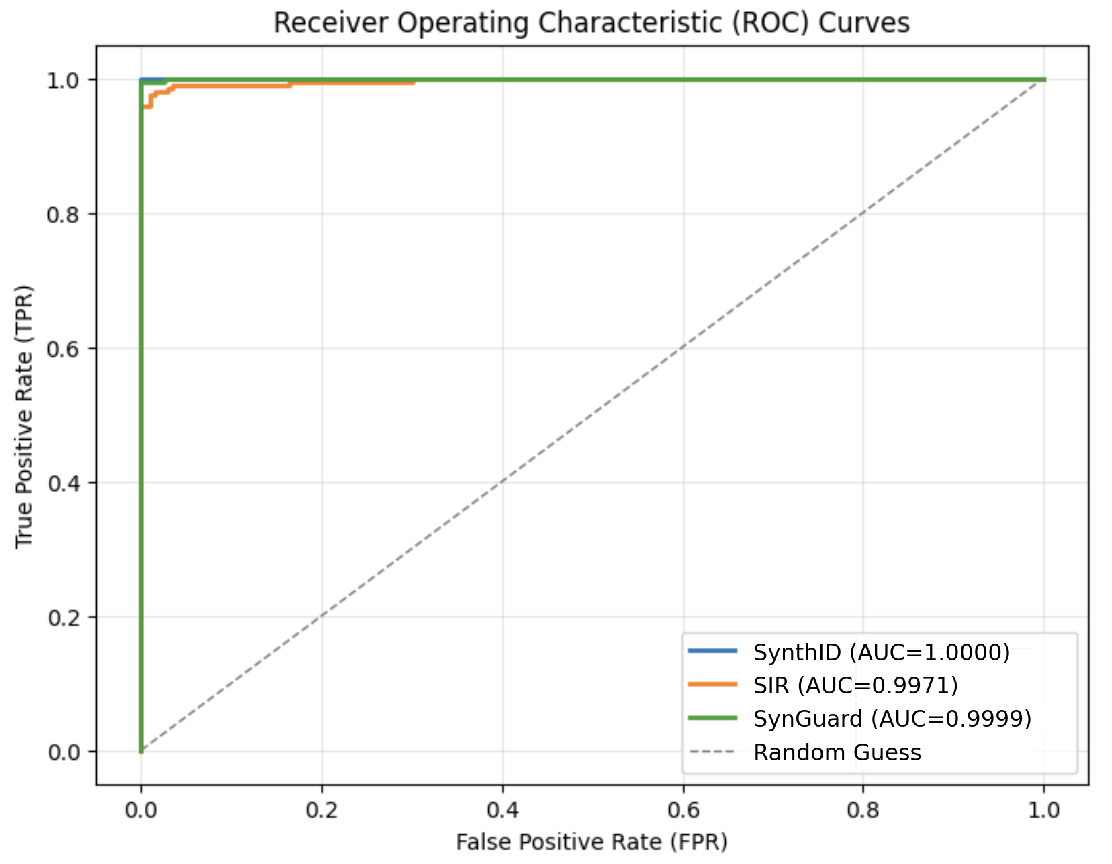}
        \caption{ROC curves of three algorithms.}
        \label{fig:roc_combined}
    \end{subfigure}
    \hspace{0.03\linewidth}
    % \hfill
    \begin{subfigure}[b]{0.47\linewidth}
        \includegraphics[width=\linewidth]{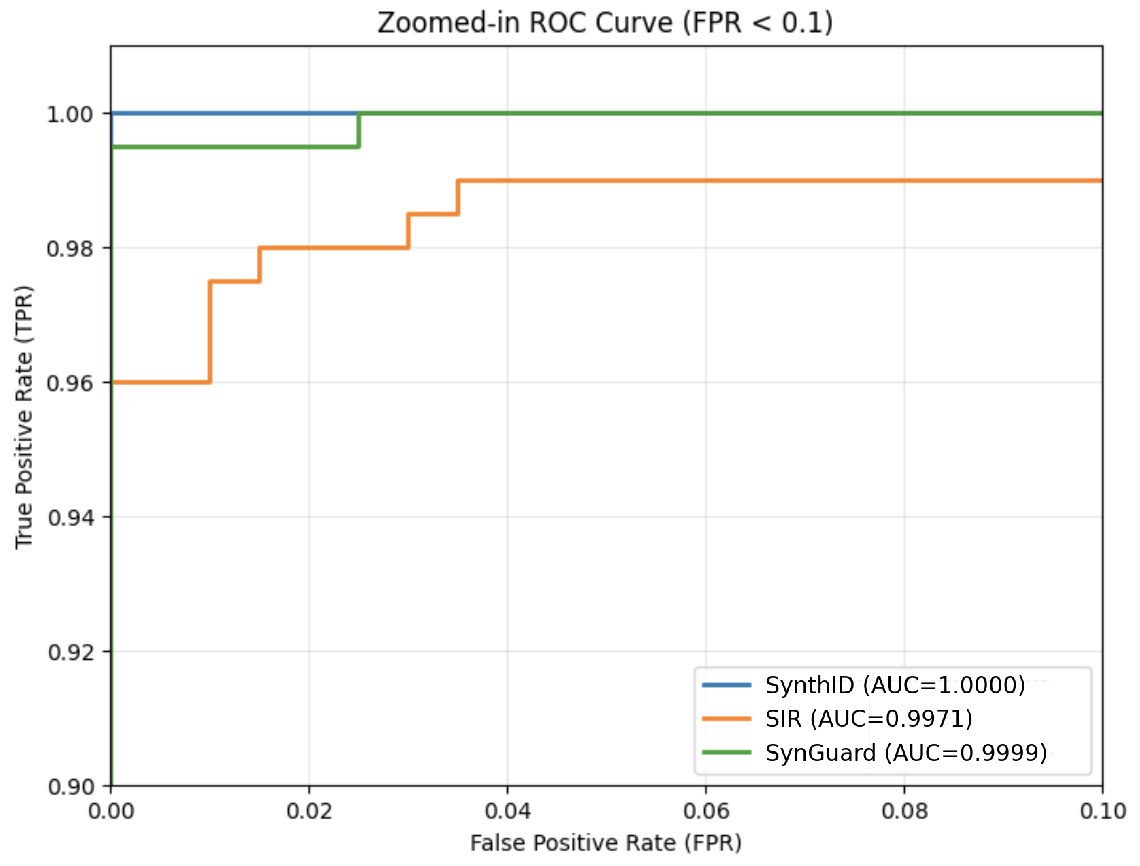}
        \caption{Zoom-in ROC curves for the three algorithms.}
        \label{fig:roc_zoomed}
    \end{subfigure}
    \caption{Comparison and zoomed-in view of ROC curves for three watermarking algorithms: SynthID-Text, SIR, and SynGuard.}
    \label{fig:roc_dual}
\end{figure}

\textbf{Detection Accuracy and ROC Curves}. Fig.~\ref{fig:roc_dual} (a) illustrates that all three algorithms achieve high detection accuracy, with AUC values above 0.9. From Fig.~\ref{fig:roc_dual} (b), it is evident that SynthID-Text achieves the highest detection accuracy of 1.00. SIR yields the lowest detection accuracy at 0.9971, exhibiting a noticeable gap compared to SynthID-Text. The detection accuracy of SynGuard is slightly lower than SynthID-Text by only 0.0001, but higher than that of SIR.

\begin{table}[tb]
    \centering
    \caption{Detection accuracy of SynthID-Text, SIR, and SynGuard.}
    \label{tab:detection accuracy comparision}
    \begin{tabular}{ccccc}     
        \toprule
        Algorithm     &  TPR   & FPR  &  \makecell{F1 with \\ best threshold}  & Running Time(s/it)\\
        \midrule
        SynthID-Text       &  1.0  &  0.0 &  1.0  & 6.09 \\
        SIR           & 0.98  & 0.015 &  0.9825 & 12.50 \\
        SynGuard   & 0.995   & 0.0 &  0.9975 & 12.93 \\
        \bottomrule
    \end{tabular}
    \vspace{-0.1in}
\end{table}

\begin{figure}[tb]
    \centering
    \includegraphics[width=0.8\linewidth]{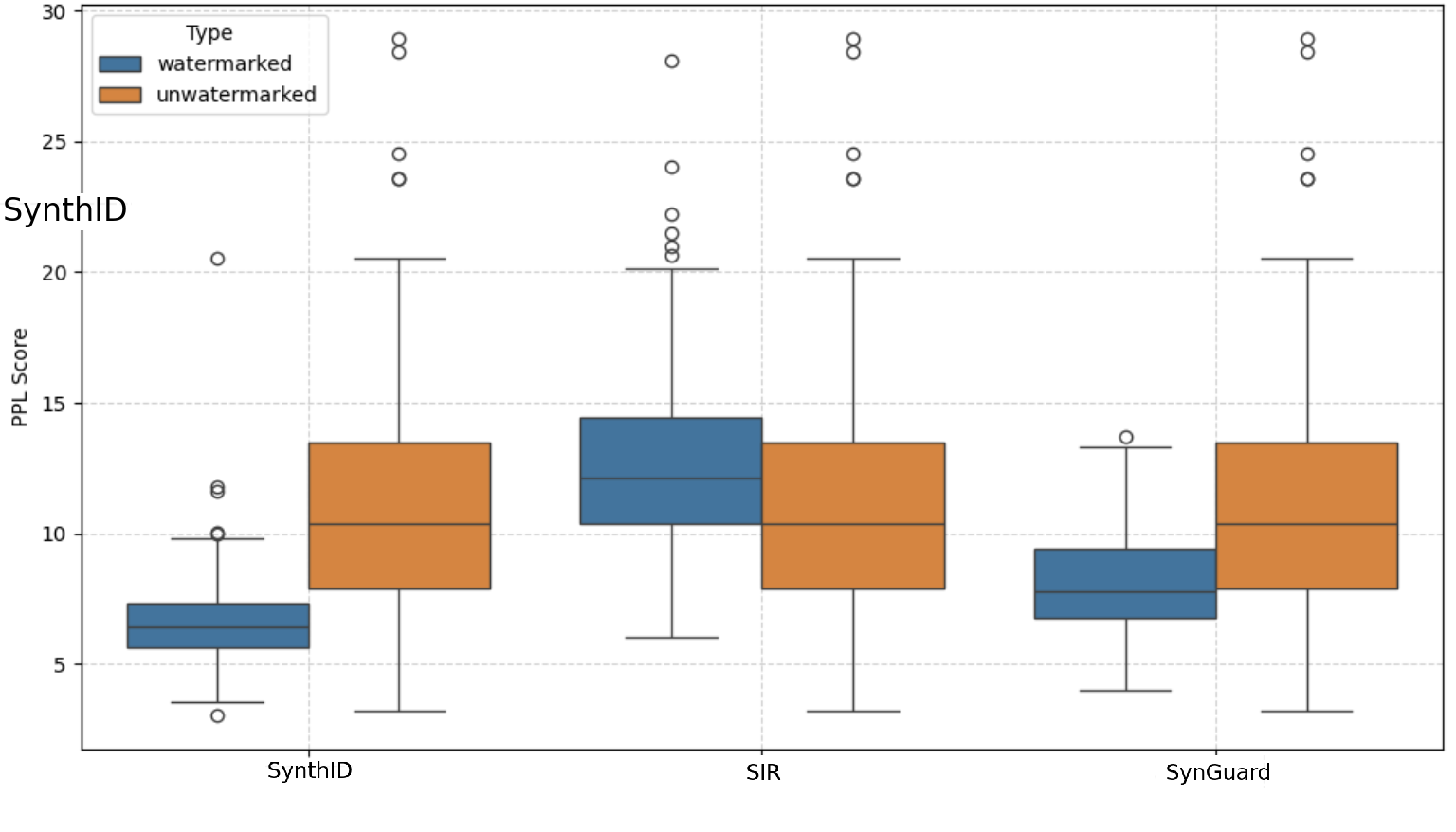}
    \caption{Text Quality Comparison Using PPL.}
    \label{fig:text quality-PPL}
    %\vspace{-0.3in}
\end{figure}

\textbf{Text Quality}. 
PPL, a metric quantifying a language model’s predictive confidence in text (lower values indicate stronger alignment with the model’s training distribution, though not absolute quality), reveals nuanced watermarking impacts in Fig. \ref{fig:text quality-PPL}. SynthID’s watermarked outputs exhibit lower PPL than their unwatermarked counterparts, suggesting its watermarking leverages semantically compatible tokens that align with the model’s learned patterns. In contrast, SIR’s watermarked texts show elevated PPL and broader distribution, indicative of disruptive interventions (e.g., forced token substitutions) that breach local coherence, amplifying predictive uncertainty. Our proposed SynGuard achieves lower PPL for watermarked texts relative to SIR, coupled with a compact distribution and minimal outliers. This arises from its hybrid design: integrating SynthID’s semantic-aware watermark encoding to preserve model-aligned fluency, while introducing stabilization mechanisms to curb output variability. Critically, PPL reflects model familiarity rather than intrinsic quality (e.g., logic or novelty), so these results underscore watermarking’s influence on textual conformity to pre-trained distributions.
% Fig. \ref{fig:text quality-PPL} compares the text quality of watermarked and unwatermarked outputs produced by these three watermarking algorithms, SynthID-Text, SIR, and SynGuard, using PPL scores as the evaluation metric. A lower PPL score indicates more natural and fluent generated text.

\textbf{Time Overhead}. Table~\ref{tab:detection accuracy comparision} reports the TPR, FPR, and F1 score for each method. The proposed SynGuard algorithm achieves an F1 score of 0.9975, just 0.25\% below the maximum value of 1. Time overhead test results are obtained from an T4 graphics card with 15.0 GB of memory on Google Colab. As can be seen, while significantly improving robustness and text quality, SynGuard did not significantly increase time overhead and is comparable to the SIR scheme. %However, its runtime is the highest among the three methods, as it combines both SIR and SynthID-Text. Notably, SIR requires approximately twice the computation time of SynthID-Text due to its use of two additional language models for semantic extraction and semantic-to-logits mapping.
% The observations are as follows:
% (1) For SynthID-Text and SynGuard, the watermarked texts exhibit lower PPL scores than their unwatermarked counterparts, indicating higher fluency and better text quality. In contrast, SIR-generated watermarked texts demonstrate higher PPL scores than unwatermarked ones.  
% (2) SynthID-Text achieves the lowest median PPL score for watermarked texts, approximately 6.5. While SynGuard yields a slightly higher median PPL around 8, this is still substantially lower than SIR’s median of 12.5.  
% (3) In particular, the box plot for SynGuard’s watermarked texts displays only one outlier, indicating consistently high quality between samples. In comparison, both SynthID-Text and SIR present multiple outliers, reflecting greater variability in their outputs.
%是衡量语言模型对文本预测能力的指标，其核心逻辑是：模型对文本的预测越 “轻松”（概率越高），PPL 值越低，反之则越高。但PPL评价的是 **“文本与模型训练分布的匹配度”**，而非直接衡量文本的 “绝对质量”（如逻辑性、相关性、信息量等）
% TODO PPL 的平均值结果
% SynthID -- PPL:{'watermarked': {'PPLCalculator': 6.577715264558792}, 'unwatermarked': {'PPLCalculator': 11.074848605394363}}
% SIR -- PPL: {'watermarked': {'PPLCalculator': 12.868765161037444}, 'unwatermarked': {'PPLCalculator': 11.074848605394363}}
% SIR_SynthID -- PPL: {'watermarked': {'PPLCalculator': 7.914364315271378}, 'unwatermarked': {'PPLCalculator': 11.074848605394363}}

\subsection{Robustness Evaluation under Attacks}
\label{6.2}

%This section presents the robustness evaluation of the proposed watermarking algorithm, SynGuard. %The attack methods and models used for testing are consistent with those described in Sessions~\ref{ch:synthID_results}.

\subsubsection{Synonym Substitution} 
\begin{figure}[tb]
    \centering
    \begin{subfigure}[b]{0.47\linewidth}
        \includegraphics[width=\linewidth]{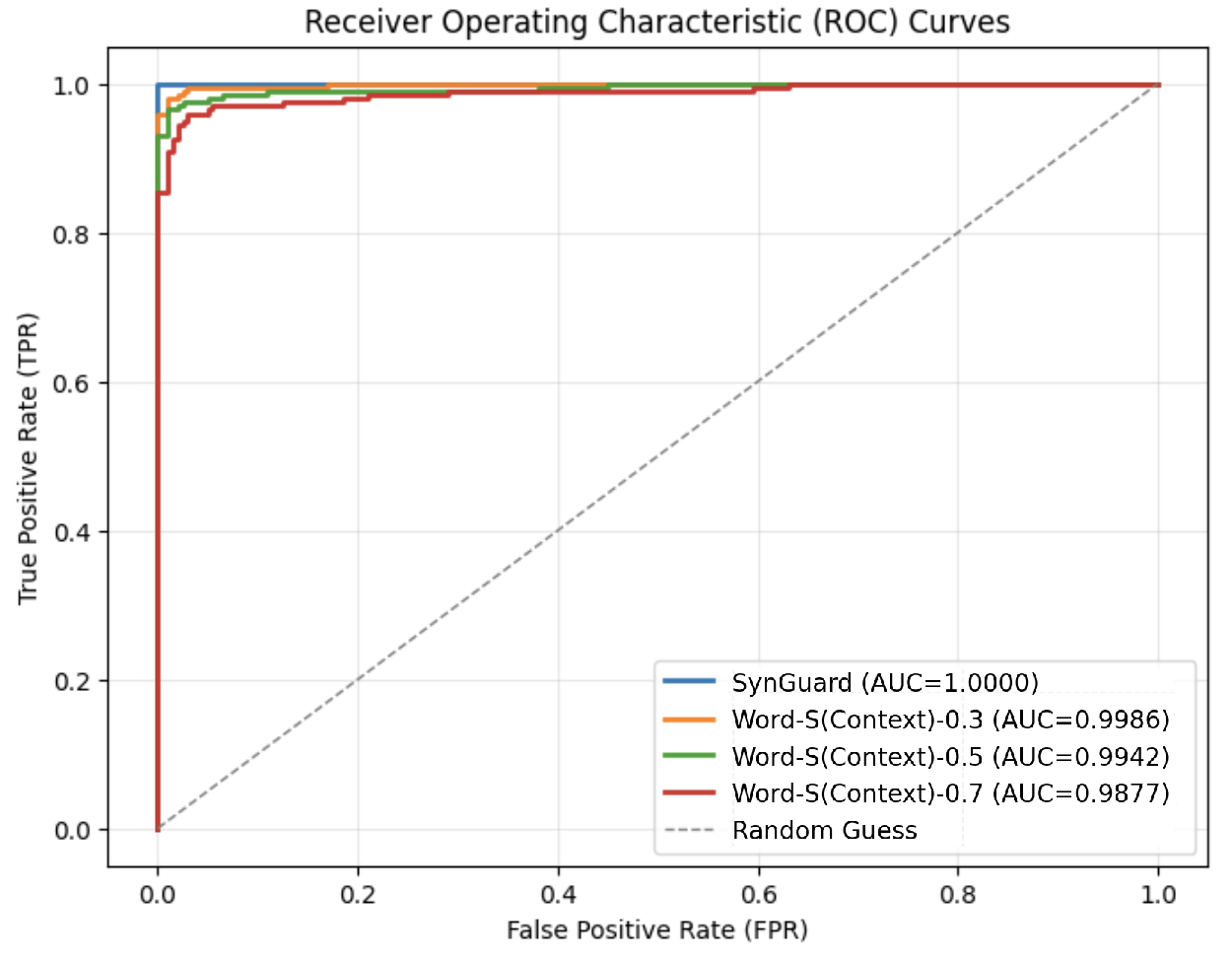}
        \caption{ROC curves}
        \label{fig:sir_synthid_synonym}
    \end{subfigure}
    \hfill
    \begin{subfigure}[b]{0.47\linewidth}
        \includegraphics[width=\linewidth]{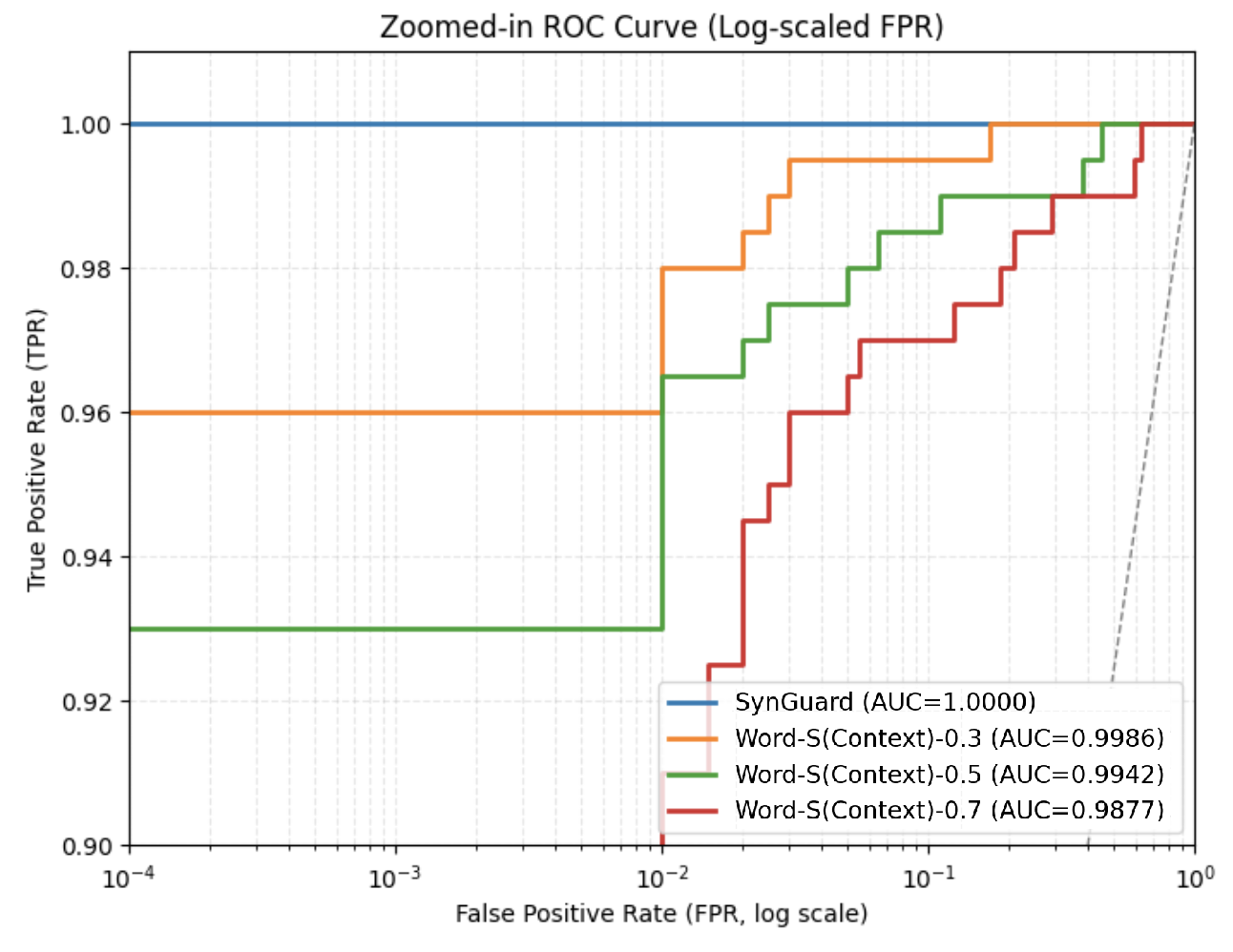}
        \caption{Zoomed-in views }
        \label{fig:new_method_zoomin}
    \end{subfigure}
    \caption{ROC curves of SynGuard under synonym substitution attacks.}
    \label{fig:roc_comparison_synonym}
    %\vspace{-0.2in}
\end{figure}

For the synonym substitution attack, we evaluated performance under varying substitution ratios: $[0, 0.3, 0.5, 0.7]$. The resulting ROC curves are shown in Fig.~\ref{fig:roc_comparison_synonym}. Even with a substitution ratio of 0.7, the AUC decreased by only 1.23\% and remained above 0.98. As shown in Table~\ref{tab:Synonym_substitution_attack_combined}, the FPR values remained low across all ratios, and the F1 scores consistently exceeded 0.95. These results highlight the strong robustness of SynGuard against synonym substitution attacks.

\begin{table}[t]
    \centering
    \caption{Watermark detection accuracy of SynthID-Text and SynGuard under different synonym substitution attacks}
    \label{tab:Synonym_substitution_attack_combined}
    \begin{tabular}{lccc|ccc}
        \toprule
        \multirow{2}{*}{\textbf{Attack}} & \multicolumn{3}{c|}{\textbf{SynthID-Text}} & \multicolumn{3}{c}{\textbf{SynGuard}} \\
        \cmidrule(lr){2-4} \cmidrule(lr){5-7}
        & TPR & FPR & F1 & TPR & FPR & F1 \\
        \midrule
        No attack              & 1.00  & 0.00  & 1.000  & 1.00  & 0.00  & 1.000 \\
        Word-S(Context)-0.3    & 0.98  & 0.005 & 0.987  & 0.98  & 0.01  & 0.985 \\
        Word-S(Context)-0.5    & 0.91  & 0.035 & 0.936  & 0.97 & 0.01  & 0.977 \\
        Word-S(Context)-0.7    & 0.82  & 0.035 & 0.884  & 0.96  & 0.03  & 0.965 \\
        \bottomrule
    \end{tabular}
\end{table}

\subsubsection{Copy-and-Paste} 
\begin{figure}[tb]
    \centering
    \begin{subfigure}[t]{0.48\linewidth}
        \centering
        \includegraphics[width=\linewidth]{figures/copy-paste_attack_roc.png}
        \caption{SynthID-Text}
        \label{fig:copy_paste_synthid}
    \end{subfigure}
    \hfill
    \begin{subfigure}[t]{0.48\linewidth}
        \centering
        \includegraphics[width=\linewidth]{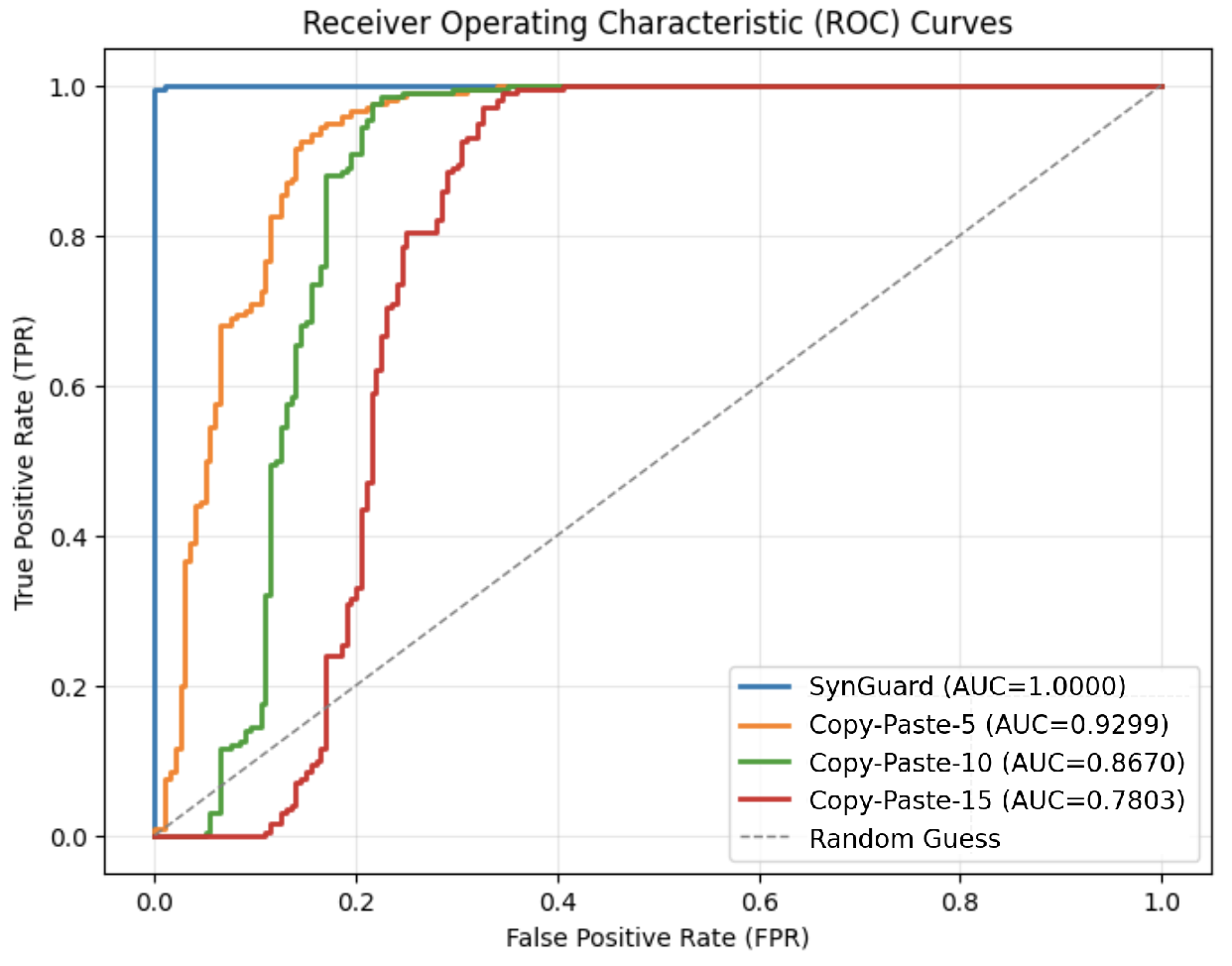}
        \caption{SynGuard}
        \label{fig:copy_paste_sir_synthid}
    \end{subfigure}
    \caption{ROC curves under different copy-and-paste attack ratios for SynthID-Text and SynGuard.}
    \label{fig:combined_copy_paste}
\end{figure}

For the copy-and-paste attack, the key parameter is the ratio between the length of the natural (or unwatermarked) text into which the watermarked content is pasted and the length of the original watermarked segment. In this experiment, the watermarked content has a fixed length of $T=200$. We tested three different length ratios: [5, 10, 15], and the results are presented in Fig.~\ref{fig:combined_copy_paste}.

Compared to synonym substitution, the impact of increasing the length ratio is more pronounced. When the copy-and-paste ratio reaches 10, the AUC already falls below 0.9. The detailed FPRs and F1 scores are listed in Table~\ref{tab:copy-and-paste_attack_combined}. Increasing the length ratio from 5 to 10 results in only a slight F1 score decrease of approximately 0.56\%. However, further increasing the ratio from 10 to 15 leads to a more substantial reduction of approximately 5\%, with the F1 score decreasing to 0.848.

% \begin{figure}[tb]
%     \centering
%     \includegraphics[width=0.8\linewidth]{figures/copy-paste attack roc.png}
%     \caption{ROC curves of SynthID-Text under different copy-and-paste attack ratios}
%     \label{fig:copy-and-paste_attack_curves}
% \end{figure}
% \begin{figure}[tb]
%     \centering
%     \includegraphics[width=0.8\linewidth]{figures/copy-and-paste_attack_curves.png}
%     \caption{ROC curves of SynGuard under different copy-and-paste attack ratios}
%     \label{fig:copy-and-paste_attack_curves}
% \end{figure}

\begin{table}[t]
    \centering
    \caption{Watermark detection accuracy under varying copy-and-paste attack settings}
    \label{tab:copy-and-paste_attack_combined}
    \begin{tabular}{lccc|ccc}
        \toprule
        \multirow{2}{*}{\textbf{Attack}} & \multicolumn{3}{c|}{\textbf{SynthID-Text}} & \multicolumn{3}{c}{\textbf{SynGuard}} \\
        \cmidrule(lr){2-4} \cmidrule(lr){5-7}
        & TPR & FPR & F1 & TPR & FPR & F1 \\
        \midrule
        No attack       & 1.0   & 0.0   & 1.0   & 1.0   & 0.0   & 1.0   \\
        Copy-Paste-5    & 0.985 & 0.245 & 0.883 & 0.95  & 0.17  & 0.896 \\
        Copy-Paste-10   & 1.0   & 0.435 & 0.821 & 0.985 & 0.225 & 0.891 \\
        Copy-Paste-15   & 0.99  & 0.485 & 0.800 & 0.99  & 0.345 & 0.848 \\
        \bottomrule
    \end{tabular}
\end{table}

% \begin{table}[tb]
%     \centering
%     \caption{Watermark detection accuracy of SynthID-Text under varying copy-and-paste attack settings}
%     \label{tab:copy-and-paste_attack_summary}
%     \begin{tabular}{lccc}     
%         \toprule
%         Attack               &  TPR   & FPR  &  F1 with best threshold \\
%         \midrule
%         No attack            &  1.0  &  0.0 &  1.0 \\
%         Copy-Paste-5     & 0.985  & 0.245 &  0.883 \\
%         Copy-Paste-10    & 1.0 & 0.435 & 0.821 \\
%         Copy-Paste-15    & 0.99   & 0.485 &  0.8 \\
%         \bottomrule
%     \end{tabular}
% \end{table}
% \begin{table}[tb]
%     \centering
%     \caption{Watermark detection accuracy of SynGuard under varying copy-and-paste attack settings}
%     \label{tab:copy-and-paste_attack_summary}
%     \begin{tabular}{lccc}     
%         \toprule
%         Attack               &  TPR   & FPR  &  F1 with best threshold \\
%         \midrule
%         No attack            &  1.0  &  0.0 &  1.0 \\
%         Copy-Paste-5     & 0.95  & 0.17 &  0.896 \\
%         Copy-Paste-10    & 0.985 & 0.225 & 0.891 \\
%         Copy-Paste-15    & 0.99   & 0.345 &  0.848 \\
%         % Copy-and-Paste-30    & 0.99  & 0.565 & 0.7750 \\
%         \bottomrule
%     \end{tabular}
% \end{table}

\subsubsection{Paraphrasing} We used the T5\footnote{https://huggingface.co/google/t5-v1\_1-xxl} model for tokenization and the Dipper\footnote{https://huggingface.co/kalpeshk2011/dipper-paraphraser-xxl} model to perform paraphrasing. The key parameters for Dipper are \texttt{lex\_diversity} and \texttt{order\_diversity}, which respectively control the lexical variation and the reordering of sentences or phrases in the generated text. 

\begin{figure}[t]
    \centering
    \begin{subfigure}[t]{0.48\linewidth}
        \centering
        \includegraphics[width=\linewidth]{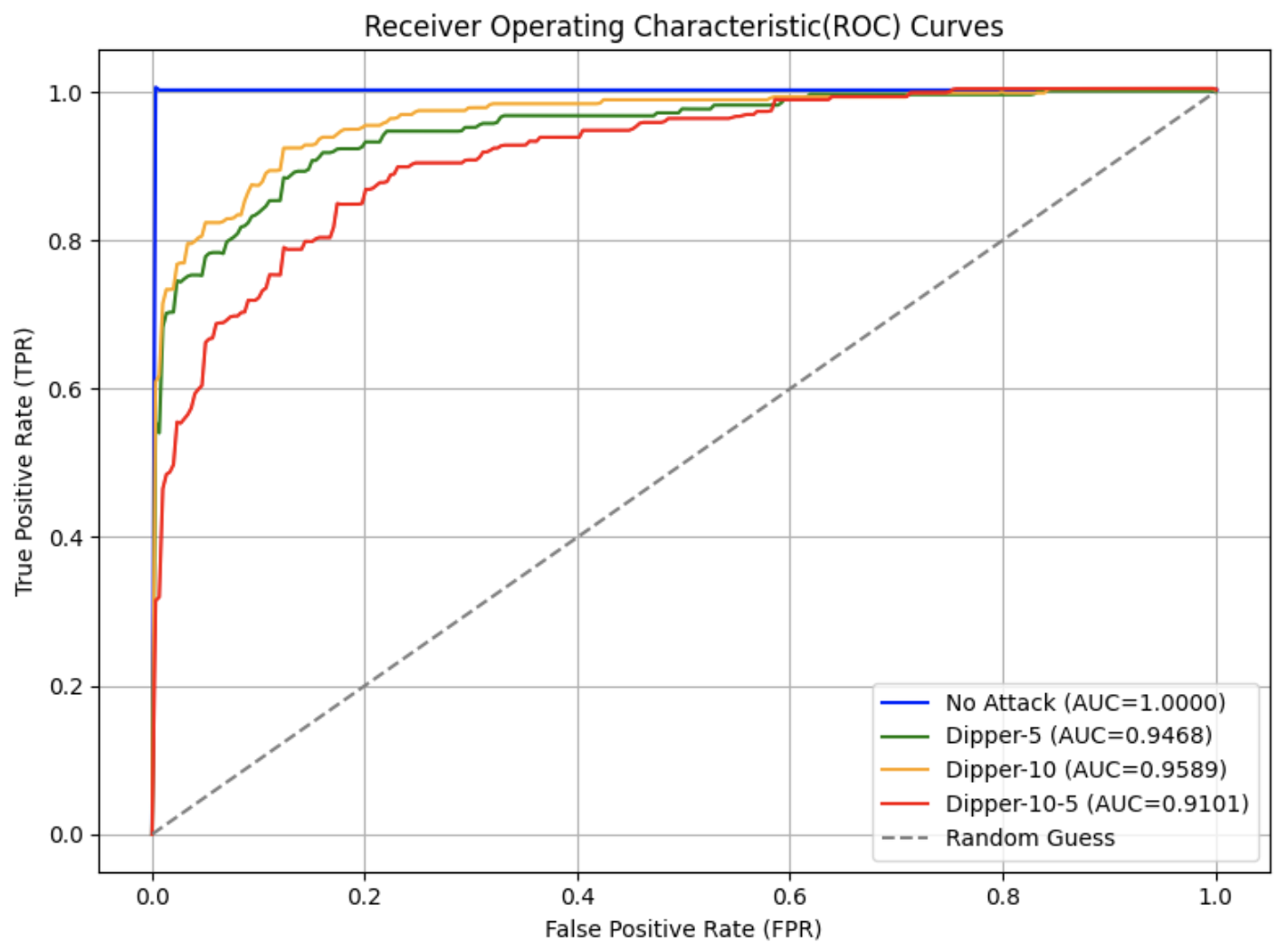}
        \caption{SynthID-Text}
        \label{fig:paraphrasing_synthid}
    \end{subfigure}
    \hfill
    \begin{subfigure}[t]{0.48\linewidth}
        \centering
        \includegraphics[width=\linewidth]{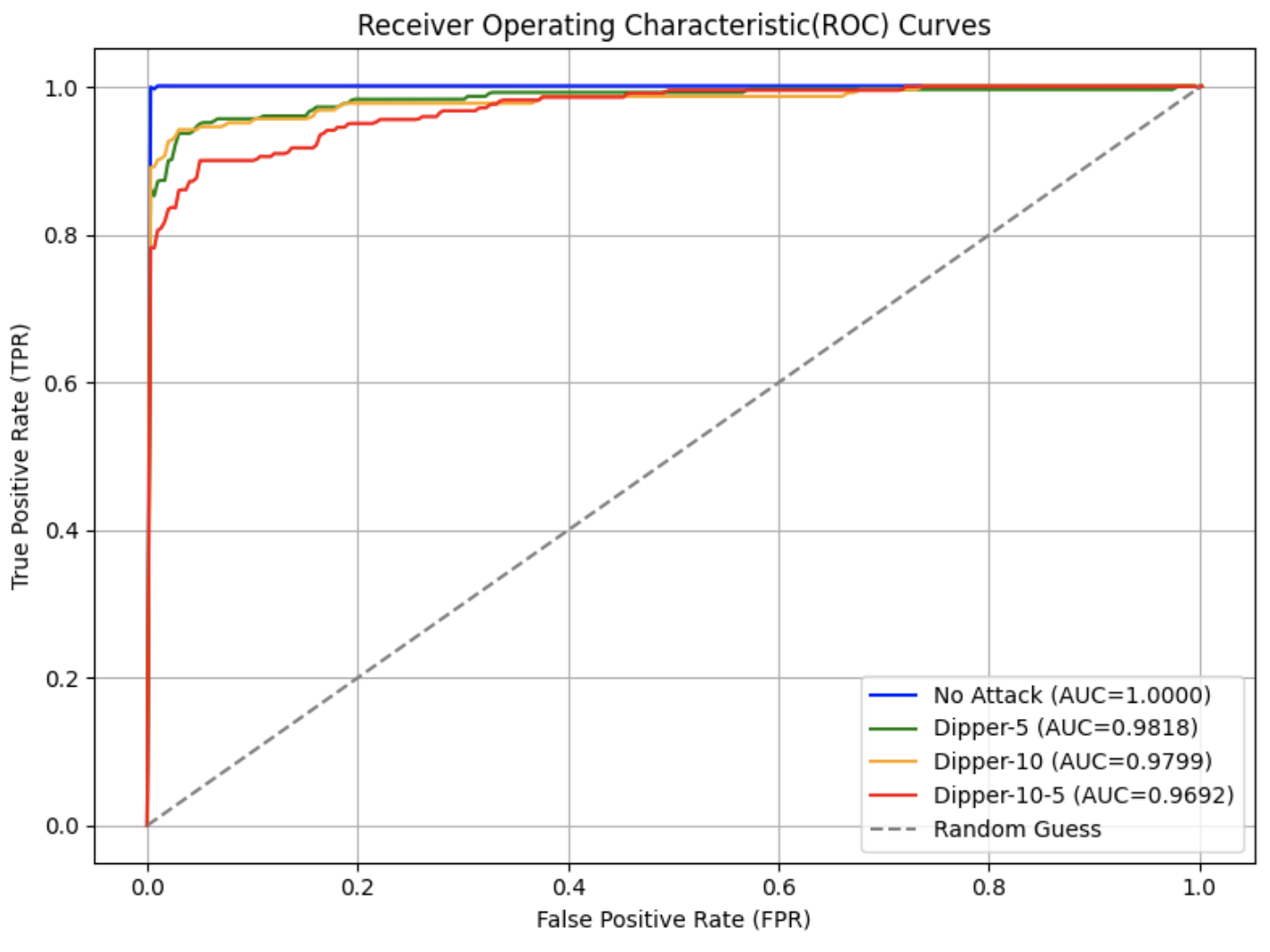}
        \caption{SynGuard}
        \label{fig:paraphrasing_sir_synthid}
    \end{subfigure}
    \caption{ROC curves under various paraphrasing attack settings for SynthID-Text and SynGuard.}
    \label{fig:combined_paraphrasing_rocs}
    %\vspace{-0.2in}
\end{figure}

In this paraphrasing attack experiment, we explored combinations of \texttt{lex\_diversity} values of 5 and 10, and \texttt{order\_diversity} values of 0 and 5. The results are shown in Fig.~\ref{fig:combined_paraphrasing_rocs}. Increasing either parameter, \texttt{lex\_diversity} or \texttt{order\_diversity}, leads to a decline in detection accuracy. 
Despite this degradation, even the most aggressive setting (\texttt{lex\_diversity} = 10 and \texttt{order\_diversity} = 5) still achieves an AUC above 0.95 and an F1 score exceeding 0.92, as reported in Table~\ref{tab:paraphrase_attacks_summary_combined}.

%下面这4张图合并成一张图
% \begin{figure}[tb]
%     \centering
%     \includegraphics[width=0.7\linewidth]{figures/rocs for paraphrasing_synthID.png}
%     \caption{ROC curves for SynthID-Text under various paraphrasing attack settings.}
%     \label{fig:combined_roc_four}
% \end{figure}
% \begin{figure}[tb]
%     \centering
%     \includegraphics[width=0.7\linewidth]{figures/rocs for paraphrasing_sir_synthID.png}
%     \caption{ROC curves for SynGuard under various paraphrasing attack settings.}
%     \label{fig:combined_roc_four}
% \end{figure}

% \begin{figure}[tb]
%     \centering
%     \begin{subfigure}[b]{0.465\linewidth}
%         \includegraphics[width=\linewidth]{figures/SynGuard-solo-roc.png}
%         \caption{No attack (baseline)}
%         \label{fig:sir_synthid_solo}
%     \end{subfigure}
%     \hfill
%     \begin{subfigure}[b]{0.465\linewidth}
%         \includegraphics[width=\linewidth]{figures/Dipper-5 new method.png}
%         \caption{ Dipper with $lex\_diversity=5$}
%         \label{fig:dipper_5}
%     \end{subfigure}

%     \vspace{0.5cm}

%     \begin{subfigure}[b]{0.465\linewidth}
%         \includegraphics[width=\linewidth]{figures/Dipper-10 new method.png}
%         \caption{ Dipper with $lex\_diversity=10$}
%         \label{fig:dipper_10}
%     \end{subfigure}
%     \hfill
%     \begin{subfigure}[b]{0.465\linewidth}
%         \includegraphics[width=\linewidth]{figures/Dipper-10-5 SynGuard.png}
%         \caption{ Dipper with  $lex\_diversity=10$ and $order\_diversity =5$}
%         \label{fig:dipper_10_5_SIR_SynthID}
%     \end{subfigure}

%     \caption{ROC curves for SynGuard under various paraphrasing attack settings.}
%      \caption*{\textbf{Note$^*$:} Due to the computational limitations of Google Colab Pro and the large sizes of the T5 and Dipper models, only one attack scenario could be executed per session. As a result, the ROC curves are presented in separate subfigures.}

%     \label{fig:combined_roc_four}
% \end{figure}

\begin{table}[t]
    \centering
    \caption{Watermark detection accuracy under different paraphrasing attack settings}
    \label{tab:paraphrase_attacks_summary_combined}
    \renewcommand{\arraystretch}{1.2}
    \begin{tabular}{lccc|ccc}
        \toprule
        \multirow{2}{*}{\textbf{Attack}} 
            & \multicolumn{3}{c|}{\textbf{SynthID-Text}} 
            & \multicolumn{3}{c}{\textbf{SynGuard}} \\
        \cmidrule(lr){2-4} \cmidrule(lr){5-7}
            & TPR & FPR & F1 & TPR & FPR & F1 \\
        \midrule
        No attack      & 1.0   & 0.0  & 1.0   & 1.0   & 0.0  & 1.0 \\
        Dipper-5       & 0.915 & 0.16 & 0.882 & 0.935 & 0.03 & 0.952 \\
        Dipper-10      & 0.92  & 0.125 & 0.900 & 0.94  & 0.03 & 0.954 \\
        Dipper-10-5    & 0.895 & 0.23 & 0.842 & 0.90  & 0.05 & 0.923 \\
        \bottomrule
    \end{tabular}
    \caption*{\textbf{Note:} \textit{Dipper-$x$} denotes the lexical diversity is $x$. \textit{Dipper-$x$-$y$} indicates lexical diversity is $x$ and order diversity is $y$.}
    \vspace{-0.15in}
\end{table}

\subsubsection{Back-translation} 
\begin{figure}[tb]
    \centering
    \begin{subfigure}[b]{0.48\linewidth}  % 第一个子图占45%行宽
        \centering
        \includegraphics[width=\linewidth]{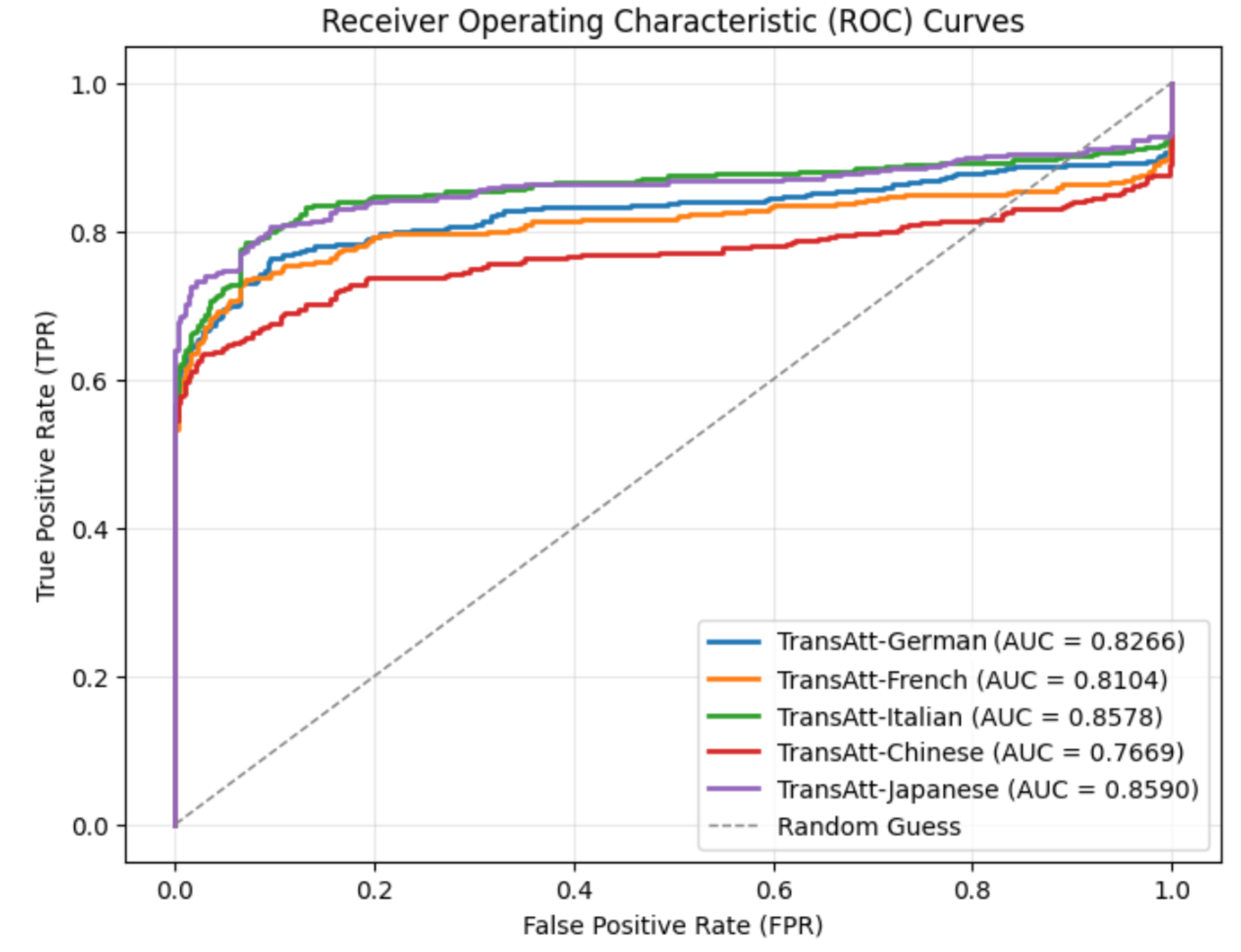}
        \caption{NLLB-200-distilled-600M}
        \label{fig:retranslation_nllb}
    \end{subfigure}
    % 移除中间的\vspace{1em}，避免上下分隔
    \hfill  % 可选：在两个子图之间添加水平填充，使间距均匀
    \begin{subfigure}[b]{0.48\linewidth}  % 第二个子图占45%行宽
        \centering
        \includegraphics[width=\linewidth]{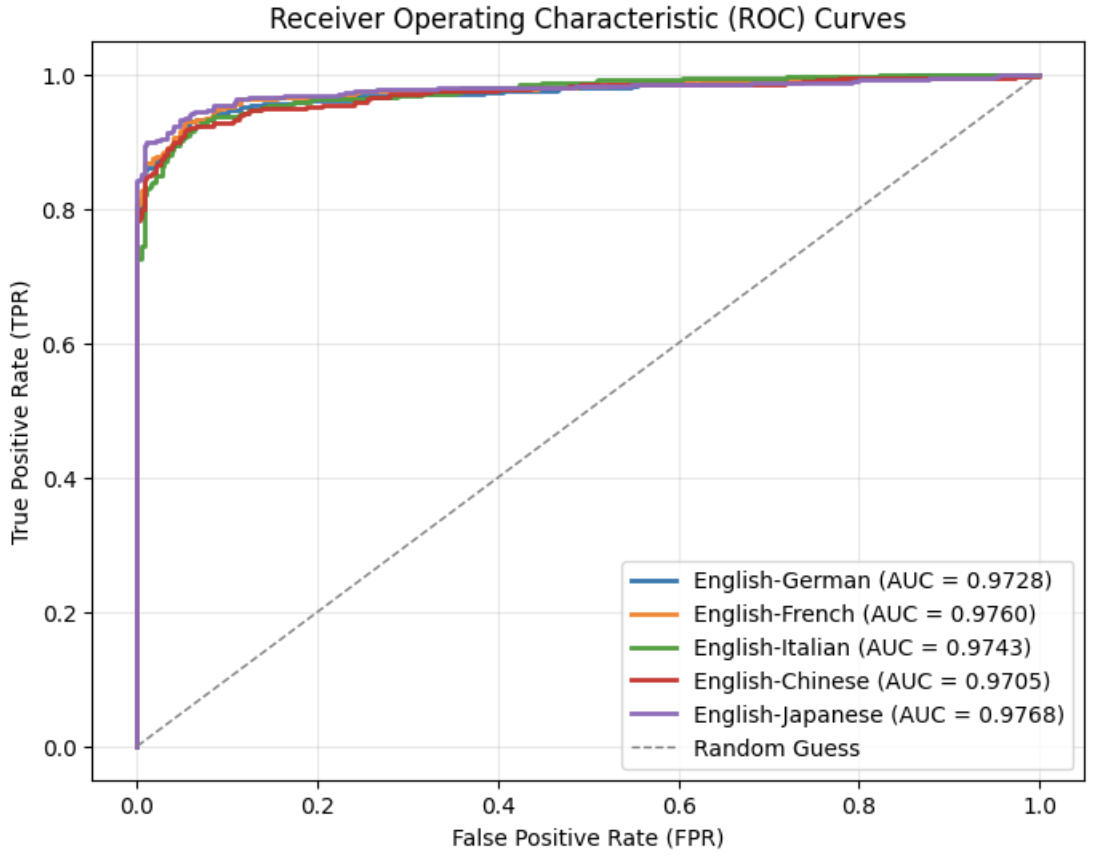}
        \caption{Google Translator}
        \label{fig:retranslation_google}
    \end{subfigure}
    
    \caption{ROC curves for back-translation on SynGuard using different translation tools.}
    \label{fig:retranslation_combined}
\end{figure}

For back-translation attack, we employed the nllb-200-distilled-600M\footnote{\url{https://huggingface.co/facebook/nllb-200-distilled-600M}} model and \texttt{googletrans} Python library to translate the original English watermarked text into different pivot languages and then back-translate it back into English. The retranslated text was subsequently used for watermark detection. The resulting ROC curves are shown in Fig.~\ref{fig:retranslation_combined}, and the results under different translators are shown in Table~\ref{tab:Retranslation_comparison_sir_synthID}. It can be observed from the results that the effectiveness of back-translation attacks is related to the translation performance of the translator for the target language, and has little to do with language-specific characteristics. Nllb is a multilingual machine translation model, with a single model handling translation for over 200 languages. In contrast, Google Translate uses dedicated machine translation models for different languages. Among the languages, back-translation attacks based on Chinese show the most significant accuracy drop and the best attack performance, which is generally consistent with the performance of machine translation. Meanwhile, the translation performance between German, French, Italian and English is better, resulting in less accuracy drop. %值得注意的是，有论文认为回译攻击的效果跟语言特性直接有关，这里我们的结论认为这种说法较为局限。我们认为：回译攻击的效果与翻译器对目标语言的翻译表现有关。语言特性决定了机器翻译模型的表现上限，而训练语料的丰富程度决定了机器翻译模型的表现上限。语言特性只是影响回译攻击的间接因素之一。

Notably, while some studies \cite{he2024can} argue that the effectiveness of back-translation attacks is directly tied to language-specific characteristics, our findings suggest this claim is rather limited. We contend that the effectiveness of back-translation attacks is instead associated with the translation performance of the translator on the target language: language-specific characteristics determine the upper bound of machine translation model performance, while the richness of the training corpus further shapes this upper bound. Consequently, language-specific characteristics constitute only one of the indirect factors influencing back-translation attacks. 
\begin{table}[tb]
    \centering
    \caption{Comparison of SynGuard watermark detection accuracy under back-translation attacks with different translation tools}
    \label{tab:Retranslation_comparison_sir_synthID}
    % \begin{tabular}{lcccccc}     
    \begin{tabular}{@{}l@{\hskip 8pt}ccc@{\hskip 6pt}|@{\hskip 3pt}ccc@{}}

        \toprule
        \multirow{2}{*}{\textbf{Attack}} & \multicolumn{3}{c}{\textbf{Nllb-200-distilled-600M}} & \multicolumn{3}{c}{\textbf{googletrans}} \\
        \cmidrule(lr){2-4} \cmidrule(lr){5-7}
        & TPR & FPR & F1 & TPR & FPR & F1 \\
        \midrule
        No attack & 0.995 & 0.0 & 0.9975 & 0.995 & 0.0 & 0.9975 \\
        Back-trans-German & 0.762 & 0.095 & 0.821 & 0.930 & 0.058 & 0.936 \\
        Back-trans-French & 0.735 & 0.070 & 0.814 & 0.930 & 0.053 & 0.938 \\
        Back-trans-Italian & 0.832 & 0.130 & 0.848 & 0.928 & 0.070 & 0.929 \\
        Back-trans-Chinese & 0.680 & 0.07 & 0.777 & 0.920 & 0.058 & 0.930 \\
        % 200 样本： 0.680 0.07 0.777
        % 400 样本： 0.702 & 0.130 & 0.766
        Back-trans-Japanese & 0.807 & 0.095 & 0.848 & 0.900 & 0.010 & 0.942 \\
        \bottomrule
    \end{tabular}
    %\vspace{-0.1in}
\end{table}

% 整合成一张图，上下排列
\captionsetup[subfigure]{labelformat=parens,labelsep=space}

\subsection{SynGuard vs. SynthID-Text}
Table~\ref{tab:sir_vs_synthid} compares SynGuard and SynthID-Text robustness under identical attacks. SynGuard achieves higher F1 scores across all evaluated attacks with the same parameters, with comparable performance in no-attack scenarios. Specifically, SynGuard retains F1 $>$ 0.9 under synonym substitution and paraphrasing, and ~0.9 under copy-and-paste, while SynthID-Text drops below 0.9 in all three. For back-translation (the most challenging attack), SynGuard outperforms SynthID-Text, with F1 rising from 0.777 to 0.711, FPR dropping from 0.225 to 0.07. Overall, F1 is improved by ~9.3\%-13\%. These results confirm SynGuard enhances detection robustness across token-level (synonym substitution), sentence-level (paraphrasing), and context-level (copy-and-paste) attacks via semantic-aware watermarking.
% Table~\ref{tab:sir_vs_synthid} compares the robustness of SynGuard and SynthID-Text under identical attack settings. Overall, SynGuard consistently achieves higher F1 scores across all evaluated attacks with the same parameters, while both methods perform comparably well in the no-attack scenario.

% Specifically, SynGuard maintains F1 scores above 0.9 under synonym substitution and paraphrasing attacks, and achieves a score close to 0.9 under the copy-and-paste attack. In contrast, SynthID-Text falls below the 0.9 threshold in all three of these scenarios.
% Regarding the most challenging attack, back-translation attack, SynGuard still outperforms SynthID-Text, improving the F1 score from 0.711 to 0.777 and reducing the FPR from 0.225 to 0.07. Overall, the F1 score has improved by approximately 9.3\% to 13\%.

% These results suggest that SynGuard improves detection robustness under a variety of attacks: token-level (e.g., synonym substitution), sentence-level (e.g., paraphrasing), and context-level (e.g., copy-and-paste), by incorporating semantic awareness into the watermarking process.

Taken collectively, our proposed SynGuard scheme exhibits computational overhead and robustness against text tampering attacks comparable to those of SIR, while demonstrating favorable text quality on par with that of SynthID-Text, thereby integrating the strengths of both approaches.
%综合来看，我们提出的SynGuard方案拥有跟SIR类似的计算开销与对文本篡改攻击的鲁棒性，又具有SynthID-Text一致的良好的文本质量，兼具了两者的优势。

\begin{table*}[t]
    \centering
    \caption{Comparison of watermark detection performance between SynGuard and SynthID-Text under various attacks }
    % scenarios}

    \label{tab:sir_vs_synthid}
    \begin{tabular}{ll|ccc|ccc}
        \toprule
        \multicolumn{2}{c|}{\textbf{Attack}} & \multicolumn{3}{c|}{\textbf{SynGuard}} & \multicolumn{3}{c}{\textbf{SynthID-Text}} \\
        Method & Parameters & TPR & FPR & F1 & TPR & FPR & F1 \\
        \midrule
        No attack & -- & 0.995 & 0.0   & \textbf{0.9975} & 1.0 & 0.0 & \textbf{1.0} \\
        \addlinespace[0.7ex]
        Substitution & $\epsilon=0.7$ & 0.96 & 0.03 & \textbf{0.965} & 0.82 & 0.035 & 0.884 \\
        \addlinespace[0.7ex]
        Copy-and-Paste & ratio=10 & 0.985 & 0.225 & 0.891 & 0.995 & \textcolor{blue}{\textbf{0.53}} & 0.788 \\
        \addlinespace[0.7ex]
        Paraphrasing & \texttt{lex}$=10$, \texttt{order}$=5$ & 0.9 & 0.05 & \textbf{0.923} & 0.895 & 0.23 & 0.842 \\
        \addlinespace[0.7ex]
        Back-Translation & \texttt{language=Chinese} & 0.680 & 0.07 & \textcolor{red}{\textbf{0.777}} & 0.675 & 0.225 & \textcolor{red}{\textbf{0.711}} \\
        % 200 SIR-SynthID：0.680  0.07 0.777
        % 200 SynthID：0.675 0.225 0.711
        \bottomrule
    \end{tabular}
    \vspace{0.5em}
    \caption*{\textbf{Note$^*$:} \textbf{Bold} F1 scores indicate values above 0.9, reflecting strong detection performance. \textcolor{blue}{Blue-highlighted} TPR or FPR values are below 0.6, suggesting performance close to random guessing. \textcolor{red}{Red-highlighted} F1 scores represent the lowest values observed across all tested attacks.}
    \vspace{-0.2in}
\end{table*}
% SynthID: using nllb-200:
% Geman: {'TPR': 0.785, 'FPR': 0.07, 'F1': 0.8463611859838275}
% French: {'TPR': 0.7075, 'FPR': 0.1725, 'F1': 0.7526595744680851}
% Italian: {'TPR': 0.675, 'FPR': 0.0425, 'F1': 0.7860262008733624}
% Chinese: {'TPR': 0.6875, 'FPR': 0.175, 'F1': 0.738255033557047}
% Japanese: {'TPR': 0.77, 'FPR': 0.0325, 'F1': 0.8543689320388349}

\subsection{Ablation Study} 
\label{5.3}

In this subsection, we investigate how the semantic weight $\delta$ affects the performance of the proposed watermarking algorithm. Based on the F1 score and AUC values from this study, we selected an optimal $\delta$ and used it for the robustness evaluations.
% under the same attack scenarios presented earlier in Section~\ref{ch:synthID_results}.

\textbf{Semantic Weight $\delta$}.  We introduce a semantic blending factor $\delta \in [0,1]$, referred to as \texttt{semantic\_weight}, to interpolate between the semantic score $s_{\text{semantic}}$ and the g-value-based score $s_{\text{g-value}}$. A larger $\delta$ emphasizes semantic coherence, while a smaller $\delta$ gives more weight to the g-value randomness statistics.

\begin{figure}[tb]
    \centering
    \begin{subfigure}[b]{0.47\linewidth}
        \includegraphics[width=\linewidth]{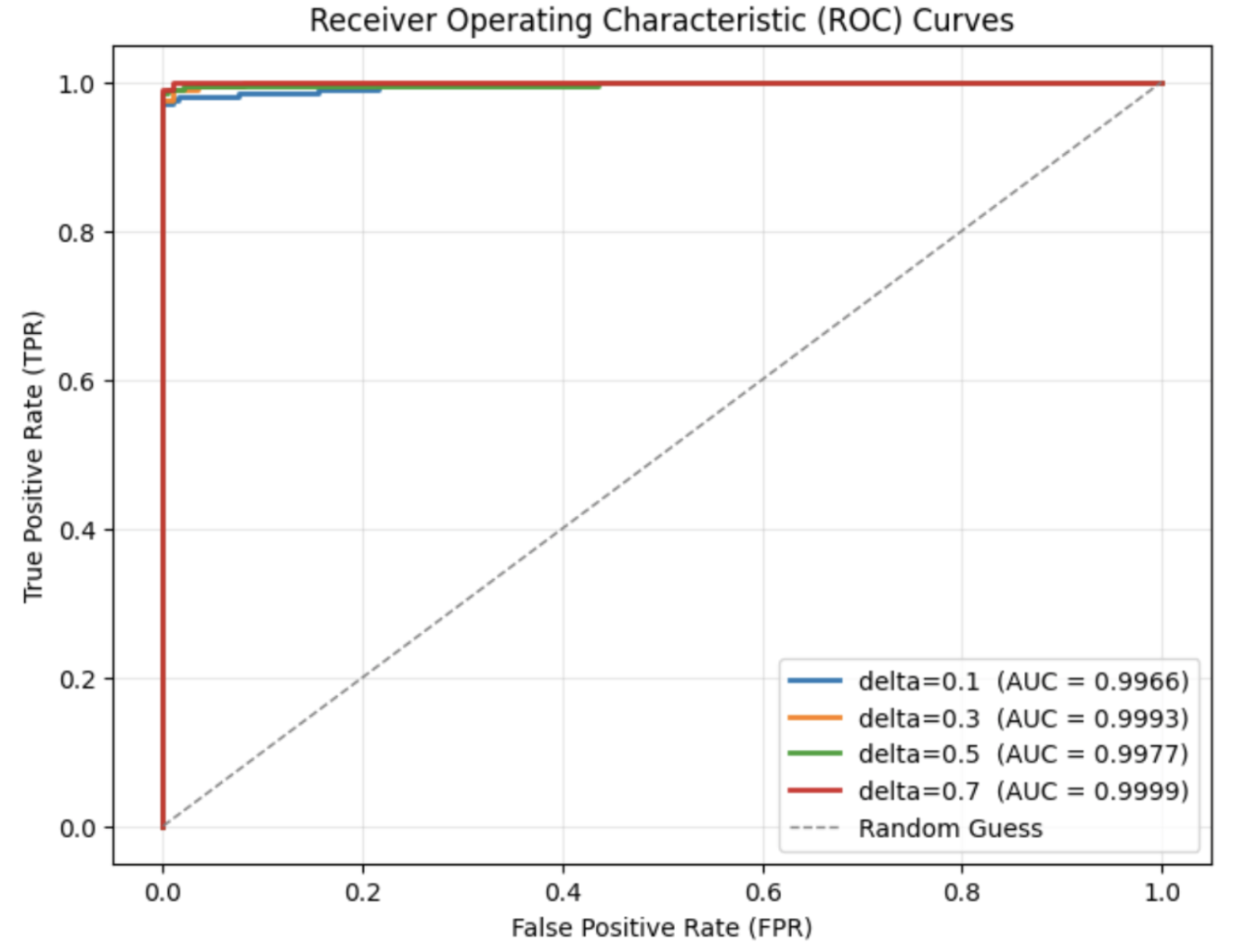}
        \caption{Regular ROC Curves}
        \label{fig:big figure}
    \end{subfigure}
    % \vspace{0.5cm}
    \begin{subfigure}[b]{0.47\linewidth}
        \includegraphics[width=\linewidth]{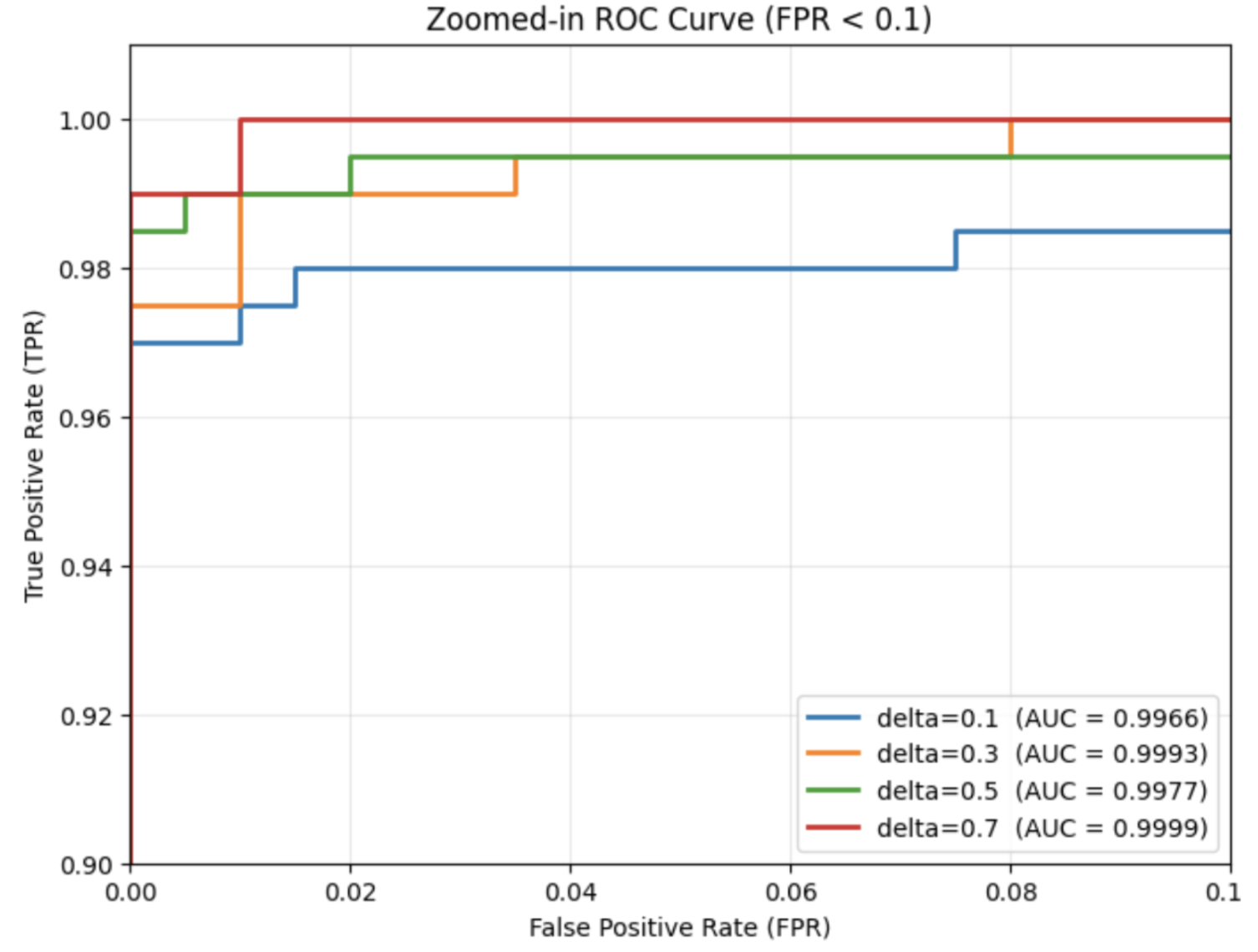}
        \caption{Zoom-in ROC Curves}
        \label{fig:zoom-in figure}
    \end{subfigure}
    \caption{ROC curves under different semantic weight settings ($\delta$)}
    % \caption{ROC curves under different $\delta$}
    \label{fig:rocs for diff delta}
    %\vspace{-0.15in}
\end{figure}

The ROC curves under different semantic weight settings are shown in Fig.~\ref{fig:rocs for diff delta}. As $\delta$ increases from 0.1 to 0.7, the AUC improves consistently. The zoomed-in view in Fig.~\ref{fig:zoom-in figure} reveals that the ROC curve for $\delta=0.7$ consistently outperforms the others.
From Table~\ref{tab:infor for diff delta}, we observe that both TPR and F1 score increase as $\delta$ grows. Although the FPR for $\delta=0.7$ is not the lowest, it is only 0.005 higher than that of $\delta=0.5$ and identical to the FPR at $\delta=0.3$. 
Therefore, in Session~\ref{ch:results for SynGuard}, we adopt $\delta=0.7$ as the default setting for the semantic weight in subsequent robustness evaluations.

\begin{table}[h!]
    \centering
    \caption{Watermark detection accuracy of SynGuard under varying semantic weights ($\delta$)}
    \label{tab:infor for diff delta}
    \begin{tabular}{cccc}     
        \toprule
        Semantic Weight $\delta$  &  TPR   & FPR  &  F1 with best threshold \\
        \midrule
        % synthID
        0.0     & 1.0    &  0    &  1.0 \\
        0.1     & 0.97    &  0    &  0.985 \\
        0.3     & 0.99    & 0.01  &  0.990 \\
        0.5     & 0.99    & 0.005 & 0.992  \\
        0.7     & 1.0     & 0.01  & 0.995  \\
        1.0     & 0.98    &  0.015    &  0.983 \\
        % sir
        \bottomrule
    \end{tabular}
    \vspace{-0.15in}
\end{table}

\section{Conclusions}
\label{ch:reflection}

This paper evaluates SynthID-Text’s robustness across diverse attacks. While SynthID-Text resists simple lexical attacks, it is vulnerable to semantic-preserving transformations like paraphrasing and back translation, which severely reduce detection accuracy. To address this, we propose SynGuard, a hybrid algorithm integrating semantic sensitivity with SynthID-Text’s probabilistic design. Via a semantic blending factor $\delta$, it balances semantic alignment and sampling randomness, boosting robustness and attack resistance. Under no-attack conditions, both methods perform comparably. For text quality, SynGuard’s slightly higher PPL score (vs. SynthID-Text) remains lower than unwatermarked text, indicating better fluency consistency. Across all attacks, SynGuard consistently outperforms SynthID-Text, improving F1 scores by 9.2\%–13\% even in pivot-language back-translation attacks (where distortion is worst). These results validate incorporating semantic information into watermarking. Overall, SynGuard is a more resilient strategy for large language models, particularly against prevalent semantic-preserving watermark removal attacks.

% \subsection{Limitations}
% While SynGuard demonstrates improved robustness under various attacks, several limitations remain that warrant further discussion.

% \textbf{External limitations } The proposed method inherits certain drawbacks from the SIR model. Specifically, it relies on multiple external language models for semantic extraction and mapping, leading to increased computational overhead and slower inference time compared to SynthID-Text. Additionally, the current implementation of SIR is limited to English, which constrains its applicability in multilingual settings.
% % As a result, SynGuard still demonstrates vulnerability to back-translation attacks involving non-English pivot languages, where semantic misalignment may disrupt watermark detection.

% \textbf{Internal limitations } The semantic blending factor $\delta$ was selected from a discrete set of candidate values, which may not include the optimal configuration. Future work could employ optimization techniques such as Bayesian Optimization or Evolutionary Algorithms to identify more precise hyperparameter. Furthermore, SynGuard currently adopts a linear combination of logits from SIR and SynthID-Text. Although effective, this approach may not fully capture the complementary strengths of the two methods. More sophisticated nonlinear fusion strategies may offer better alignment and resilience, representing a promising direction for future research.

% \section*{References}
% \bibliography{references}
\bibliographystyle{IEEEtran}
\bibliography{references}  % 不需要加 .bib 后缀

% Please number citations consecutively within brackets \cite{b1}. The 
% sentence punctuation follows the bracket \cite{b2}. Refer simply to the reference 
% number, as in \cite{b3}---do not use ``Ref. \cite{b3}'' or ``reference \cite{b3}'' except at 
% the beginning of a sentence: ``Reference \cite{b3} was the first $\ldots$''

% Number footnotes separately in superscripts. Place the actual footnote at 
% the bottom of the column in which it was cited. Do not put footnotes in the 
% abstract or reference list. Use letters for table footnotes.

% Unless there are six authors or more give all authors' names; do not use 
% ``et al.''. Papers that have not been published, even if they have been 
% submitted for publication, should be cited as ``unpublished'' \cite{b4}. Papers 
% that have been accepted for publication should be cited as ``in press'' \cite{b5}. 
% Capitalize only the first word in a paper title, except for proper nouns and 
% element symbols.

% For papers published in translation journals, please give the English 
% citation first, followed by the original foreign-language citation \cite{b6}.

\end{document}